\documentclass[10pt]{article}

\usepackage{amsmath,amssymb}

\usepackage{changepage}

\usepackage[utf8x]{inputenc}

\usepackage{textcomp,marvosym}

\usepackage{cite}

\usepackage{nameref,hyperref}

\usepackage[right]{lineno}

\usepackage{microtype}
\DisableLigatures[f]{encoding = *, family = * }

\usepackage[table]{xcolor}

\usepackage{array}

\newcolumntype{+}{!{\vrule width 2pt}}

\newlength\savedwidth

\newcommand{\PP}{{\rm P}}
\newcommand{\mc}{\mathcal}
\newcommand{\ep}{\epsilon}

\renewcommand{\d}{{\rm d}}
\newcommand{\e}{{\rm e}}
\newcommand{\pd}{\partial}

\usepackage[aboveskip=1pt,labelfont=bf,labelsep=period,justification=raggedright,singlelinecheck=off]{caption}

\makeatletter
\renewcommand{\@biblabel}[1]{\quad#1.}
\makeatother

\date{}

\usepackage{lastpage,fancyhdr,graphicx}
\usepackage{epstopdf}

\begin{document}
\vspace*{0.2in}

\begin{flushleft}

{\Large \bf Synaptic mechanisms of interference in working memory
}
\newline
\\
Zachary P Kilpatrick\textsuperscript{1,2}
\\
\bigskip
\textbf{1} Department of Applied Mathematics, University of Colorado, Boulder CO, USA
\\
\textbf{2} Department of Physiology and Biophysics, University of Colorado School of Medicine, Aurora CO, USA
\\

%
%




\textbf{*} zpkilpat@colorado.edu
\end{flushleft}
\section*{Summary}
Information from preceding trials of cognitive tasks can bias performance in the current trial, a phenomenon referred to as interference. Subjects performing visual working memory tasks exhibit interference in their trial-to-trial response correlations: the recalled target location in the current trial is biased in the direction of the target presented on the previous trial. We present modeling work that (a) develops a probabilistic inference model of this history-dependent bias, and (b) links our probabilistic model to computations of a recurrent network wherein short-term facilitation accounts for the dynamics of the observed bias. Network connectivity is reshaped dynamically during each trial, providing a mechanism for generating predictions from prior trial observations. Applying timescale separation methods, we can obtain a low-dimensional description of the trial-to-trial bias based on the history of target locations. The model has response statistics whose mean is centered at the true target location across many trials, typical of such visual working memory tasks. Furthermore, we demonstrate task protocols for which the plastic model performs better than a model with static connectivity: repetitively presented targets are better retained in working memory than targets drawn from uncorrelated sequences. \\
\vspace{-2mm}

\noindent
{\bf Keywords:} working memory; interference; short-term facilitation; probabilistic inference; bump attractor



\section*{Introduction}

Parametric working memory experiments are a testbed for behavioral biases and errors, and help identify neural mechanisms that underlie them~\cite{funahashi89,romo99}. In visuospatial working memory, subjects identify, store, and recall target locations in trials lasting a few seconds. Response errors are normally distributed~\cite{white94,wimmer14}, and tend to accumulate during the delay-period, while subjects retain the target location in memory~\cite{funahashi89,wimmer14}.
Complementary neural recordings suggest these working memories are implemented in circuits comprised of stimulus-tuned neurons with slow excitation and broad inhibition~\cite{goldmanrakic95,compte00}. Persistent activity emerges as a tuned pattern of activity called a bump state, whose peak encodes the remembered target position~\cite{renart03,wimmer14}.

Neuronal studies of visual working memory typically focus on population activity within a single trial, ignoring serial correlations across trials~\cite{constantinidis16}. Several authors have identified behavioral biases that cause the previous trial's visual target to interfere with the subject's response on the subsequent trial~\cite{papadimitriou15}. For instance, in delayed match-to-sample tests, false alarms occur more often when comparison stimuli match samples from previous trials~\cite{makovski08}. Originally, interference was observed in verbal working memory tasks~\cite{keppel62}, and evidence suggests the effect impacts working memory capacity~\cite{kane00}. One consistent observation is that interference is reduced by increasing the time interval between trials~\cite{dunnett90,papadimitriou15}, suggesting the effect persists for a few seconds. Investigations of interference in visuospatial working memory reveal other effects: Increasing the delay-period of working memory trials increases the bias strength, and responses are biased in the direction of the stimulus from the previous trial~\cite{papadimitriou15}.


Our study focuses on why and how interference biases arise visuospatial working memory. First, what evidence accumulation strategy accounts for the bias introduced by the previous trial's target? We will show these biases emerge in observers using sequential Bayesian updating to predict the location of the next target.
Such models are obtained by iteratively applying Bayes' rule to a stream of noisy measurements, updating an observer's belief of the most likely choice.
In changing environments, older measurements are discounted at a rate that increases with the assumed change rate of the environment~\cite{glaze15,velizcuba16}.
In our model, the sequence of targets observed on each trial is used to predict the next target. When subjects assume the environment changes rapidly, only the most recent target is used to make their prediction, leading to suboptimal inference of the subsequent target~\cite{beck12}.


What neurophysiological processes could account for intertrial biases? Both the decay and activation timescales of the bias appear to be on the order of seconds.
We propose short-term facilitation (STF), which acts on the timescale of seconds~\cite{tsodyks97}, can account for the dynamics of the bias.
In a recurrent network that sustains persistent activity during a delay-period in the form of an activity bump, facilitated synapses from neurons tuned to the previous target attract the activity bump in the subsequent trial. Previous models identified STF as a possible mechanism for lengthening the timescale of working memory~\cite{mongillo08,itskov11,mi17}. Our study proposes interference arises as a result of an irrelevant working memory remaining from the previous trial.


Our neurocomputational model accounts for recent observations of interference from visual working memory experiments~\cite{papadimitriou15}, and makes novel predictions linking behavioral responses to corresponding neural and synaptic mechanisms. The separation in timescales between the neural activity dynamics and STF variable allows us to derive a low-dimensional model describing the bump's interaction with the network's evolving synaptic weights. We find that protocols with a uniform distribution of possible target angles lead to response error distributions that are normally distributed about zero as found previously~\cite{funahashi89,wimmer14,constantinidis16}. Conversely, target protocol sequences with strong serial correlations can lead to a biased distributions in recalled target positions. Such biases may be advantageous in more complex tasks, where information from previous trials provides information about the target location in subsequent trials, as we show. Finally, we demonstrate that a recurrent network with STF supports bump attractors whose diffusion time course possesses two distinct phases, a prediction we propose to validate our model.

\section*{Results}

\begin{figure}[hbtp]
\begin{center} \includegraphics[width=15cm]{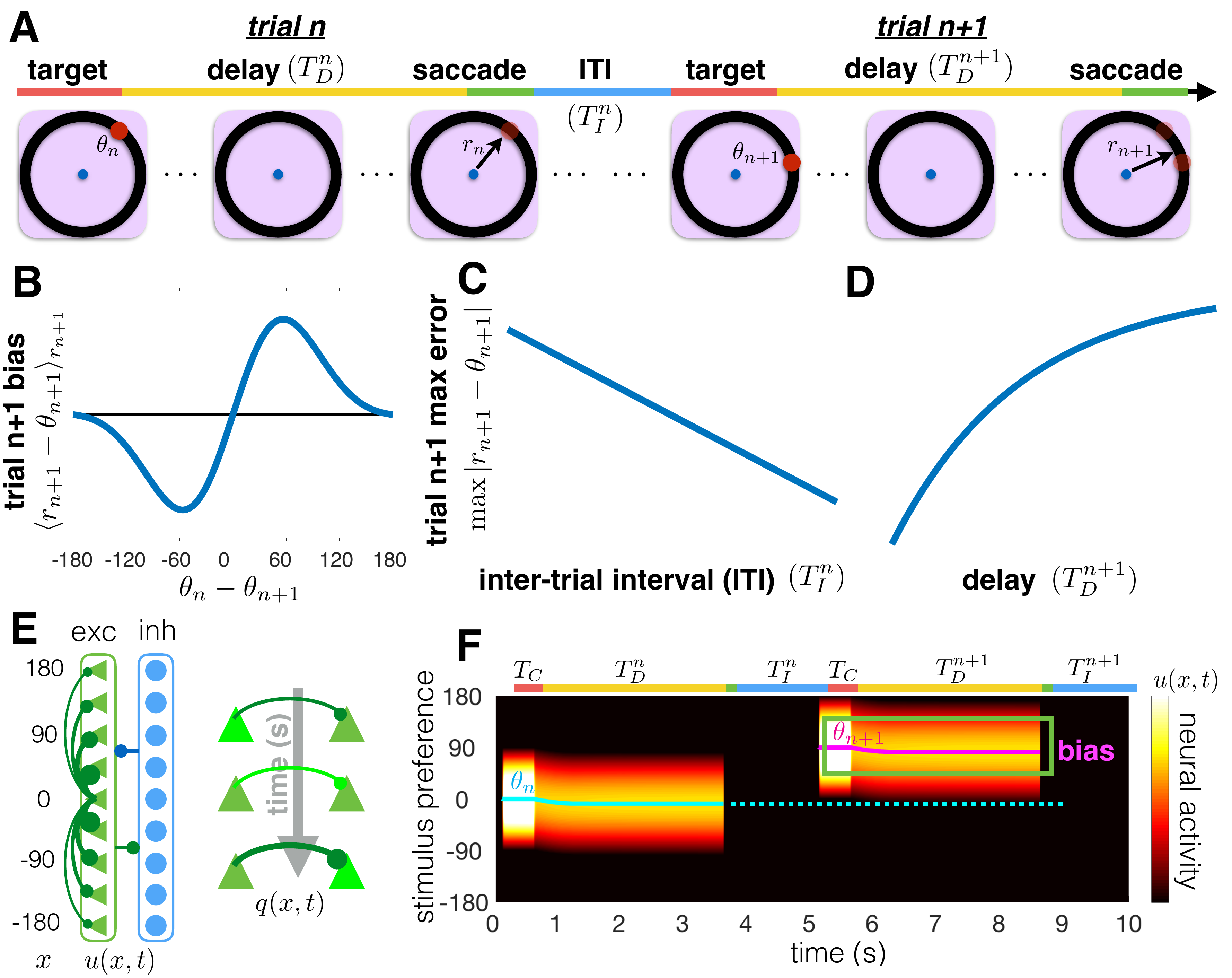} \end{center}
\caption{Interference in visuospatial working memory, and our corresponding recurrent network model with STF. (A) A visuospatial working memory task was administered in consecutive trials, adapted from Papadimitriou et al. (2015)~\cite{papadimitriou15}. The subject fixates on the central (blue) dot and a target (red dot) appears at location $\theta_n$ on trial $n$. After the target disappears, the subject retains a memory of the target location during the delay-period ($T_D^n$ and $T_D^{n+1}$, 0-6000ms). Lastly, the subject makes a saccade ($r_n$ and $r_{n+1}$) to the remembered target location. Papadimitriou et al. (2015) found a systematic impact of the relative location ($\theta_n - \theta_{n+1}$) of the trial $n$ target on the trial $n+1$ response $r_{n+1}$. (B) Response errors in trial $n+1$ ($\langle r_{n+1} - \theta_{n+1} \rangle_{\theta_{n+1}}$) depend on the relative location of the target ($\theta_n - \theta_{n+1}$) in trial $n$. Responses err in the direction of the previous target $\theta_n$, but this tendency is non-monotonic in $\theta_n - \theta_{n+1}$. (C,D) The maximum average error in trial $n+1$ decreases with intertrial interval $T_I^n$ (panel C) and increases with the trial $n+1$ delay-period $T_D^{n+1}$ (panel D). (E) Schematic of our recurrent network model, showing excitatory (triangle) and inhibitory (circles) neurons. Connections between excitatory cells are distance-dependent. Effects of the inhibitory population are fast and spatially uniform, so excitatory and inhibitory populations are merged into single variable $u(x,t)$. STF increases the strength of recently used synapses, described by the variable $q(x, t)$. (F) A tuned input during the cue period ($T_C$) generates a bump of neural activity $u(x,t)$ centered at $x=\theta_n$ that persists during the delay-period of trial $n$ ($T_D^n$) and ceases after the response. After the intertrial interval ($T_I^n$), the bump initially centered at $x=\theta_{n+1}$ drifts towards the position of the bump in the previous trial (dotted line) due to the attractive force of STF. Input fluctuations are ignored here to highlight the bias in a single trial.}
\label{fig1_task}
\end{figure}

Our study presents two frameworks for generating interference in a sequence of visual working memory trials. Both models use information about the target location on the previous trial to bias the response on the current trial. First, we develop a probabilistic inference model that predicts the distribution of possible target angles on the current trial based on observations of past trials. When the observer assumes the environment changes rapidly, the predictive distribution is primarily shaped by the previous trial's target. Second, we analyze a recurrent network model with STF wherein a localized bump of activity represents the observer's belief on the current trial and the spatial profile of STF represents the observer's evolving predictive distribution for the subsequent target.
We show the attractor structure of the network model can be directly related to the predictive distribution of the inference model.

\subsection*{Interference in a visual working memory task}

We focus specifically on an oculomotor delayed-response task with a single target presented in each trial~\cite{funahashi89,wimmer14}.
On each trial, the subject views a target $\theta_n$ during a short cue period (Fig. \ref{fig1_task}A). They must remember the target location during a delay-period and saccade to the remembered location at the end. Response errors depend on the previous trial in three distinct ways, originally reported in Papadimitriou et al. (2015)~\cite{papadimitriou15}: (1) responses are attracted to the location of the previous target, graded with the difference between the current and previous target (Fig. \ref{fig1_task}B); (2) the bias decreases as the interval between trials is increased (Fig. \ref{fig1_task}C); and (3) the bias increases as the delay-period increases (Fig. \ref{fig1_task}D). As we will show, these biases are captured by a model of an observer that predicts the current target based on the previous target. These effects also emerge in a recurrent network model with slow excitation, subject to STF, and broad inhibition (Fig. \ref{fig1_task}E). This network represents the memory of the presented target as a bump of neural activity, which drifts in the direction of the target presented on the previous trial (Fig. \ref{fig1_task}F). Before analyzing the mechanics of this network model in more detail, we derive a probabilistic inference model that accounts for these tendencies.

\subsection*{Inference model for updating target predictions}

\begin{figure}[t]
\vspace{-7mm}
\begin{center} \includegraphics[width=15cm]{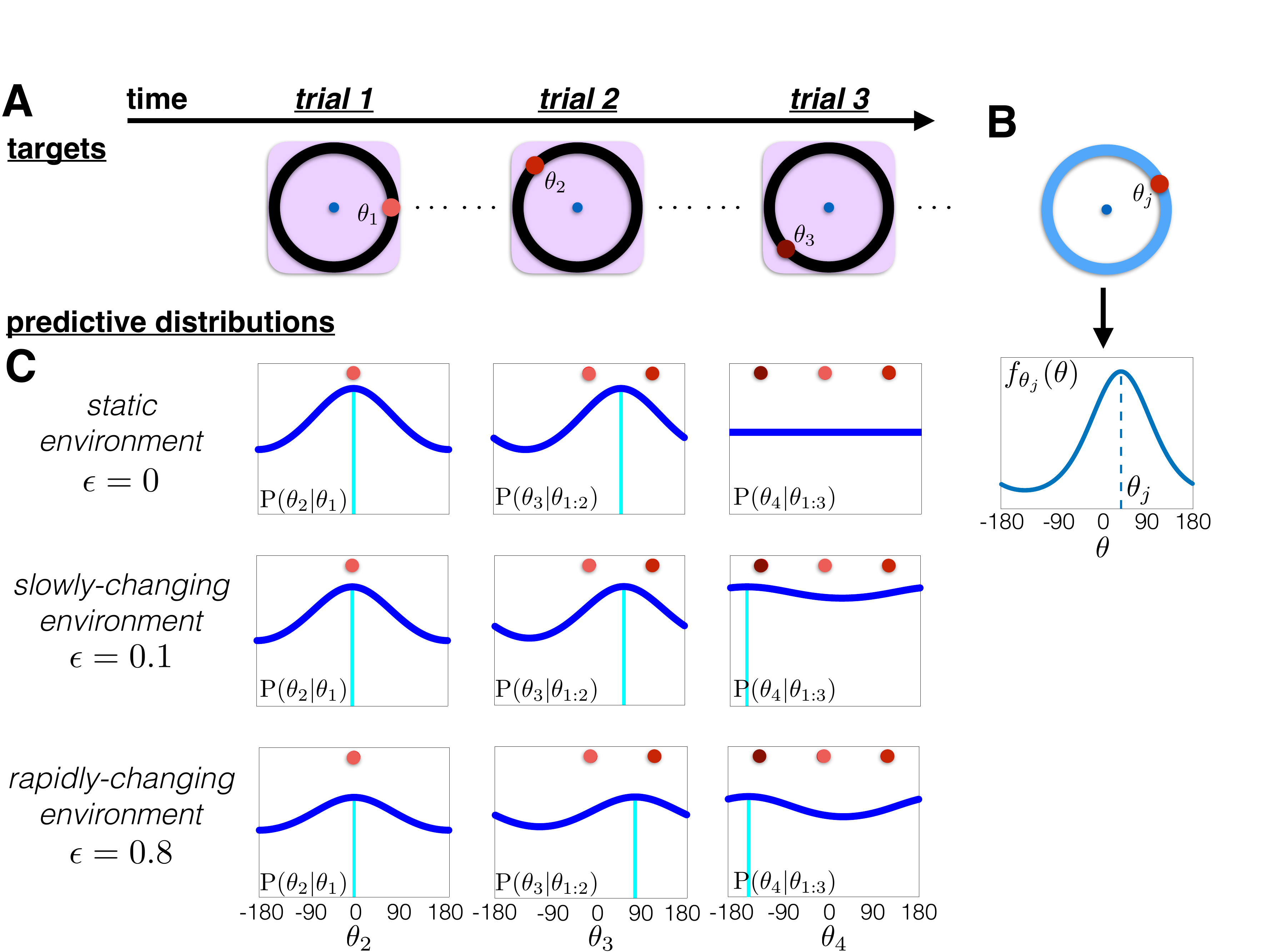} \end{center}
\caption{Updating the predictive distribution $L_{n+1,\theta}$. The observer infers the predictive distribution for the subsequent target $\theta_{n+1}$ from prior observations $\theta_{1:n}$, assuming a specific change rate $\ep$ of the environment: $L_{n+1,\theta} : = \PP(\theta_{n+1} = \theta|\theta_{1:n}, \ep)$. (A) A sequence of presented targets: $\theta_{1:3}$. Note the environment is typically static, so the true $\ep_{\text{true}} =0$. (B) Probability $f_{\theta_j} (\theta)$, peaked and centered at $\theta_j$, showing the observer's assumed probability that $\theta_{n+1} = \theta$, if $\theta_j$ is observed on trial $j$ and the distribution remains the same in between ($s_{n+1}(\theta) \equiv s_j(\theta)$). (C) Evolution of the predictive distribution $L_{n+1,\theta} : = \PP(\theta_{n+1} = \theta |\theta_{1:n}, \ep)$ for static ($\ep = 0$); slowly-changing ($\ep= 0.1$); and rapidly-changing ($\ep =0.8$) environments. In static environments, all observations $\theta_{1:3}$ are weighted equally whereas in the rapidly-changing environment, the most recent observation dominates.}
\label{fig2_preddist}
\end{figure}

Our goal now is to show that sequential Bayesian updating can account for interference observed in working memory, given specific constraints on a probabilistic updating algorithm. The observer attempts to predict the probability of observing target angle $\theta_{n+1} = \theta$ in trial $n+1$, given the targets $\theta_{1:n} = \{ \theta_1, \theta_2, ..., \theta_n\}$ observed in the previous $n$ trials (Fig. \ref{fig2_preddist}A).
However, the target $\theta_j$ on the $j^{\text{th}}$ trial will only help predict the target $\theta_{n+1}$ on the $n+1^{\text{th}}$ trial if the distribution $s_{n+1}(\theta)$ from which targets are drawn remains the same between trial $j$ and trial $n+1$~\cite{wilson10}.
The observer assume the distribution from which presented targets are drawn changes stochastically at a fixed rate $\ep : = \PP (s_{n+1}(\theta) \not \equiv s_n(\theta))$.
Most visual working memory protocols fix the distribution of target angles throughout the task ($\ep = 0$)~\cite{funahashi89,pesaran02,wimmer14,papadimitriou15}, as we do for most of our study, so the observer employs a potentially incorrect model to estimate this distribution ($\ep >0$). Subjects in psychophysical tasks can have a strong bias toward assuming environments change on a timescale of several seconds~\cite{glaze15}, and this bias is not easily trained away~\cite{beck12,navarro14}. Combining these features of the model, the observer updates their predictive distribution for the target during the $(n+1)^{\text{th}}$ trial.


Our
algorithm is based on models that compute a predictive distribution for a stochastically moving target, given a sequence of noisy observations~\cite{adams07,wilson10}.
The predictive distribution is computed using sequential analysis~\cite{wald48,velizcuba16}: Prior to trial $n+1$, the observer has seen $n$ targets $\theta_{1:n} = \{ \theta_1, \theta_2, ..., \theta_n \}$.
The observer computes $f_{\theta'}(\theta): = \PP( \theta_{n+1} = \theta | \theta_j = \theta' , s_{n+1}(\theta) \equiv s_j(\theta))$ (Fig. \ref{fig2_preddist}B), the probability of observing the target $\theta_{n+1}$ in the $(n+1)^{\text{th}}$ trial assuming
the underlying probability distribution from which targets are sampled does not change from trial $j$ to $n+1$ ($s_{n+1}(\theta) \equiv s_j(\theta)$), for each trial $j=1,...,n$. The true distribution of target angles $\theta$ remains uniform throughout most our study, so the observer is applying suboptimal inference. Further details of our Bayesian nonparametric model are given in Methods.

The observer thus computes a predictive distribution $L_{n+1,\theta} = \PP( \theta_{n+1} = \theta | \theta_{1:n}, \ep)$, using the previous targets $\theta_{1:n}$ (Fig. \ref{fig2_preddist}A) to predict the subsequent target $\theta_{n+1}$. If the observer assumes the distribution $s_{n+1}(\theta)$ from which targets are drawn in trial $n+1$ changes stochastically with a rate $\ep \in (0,1)$, recent observations will be weighted more in determining $L_{n+1,\theta}$~\cite{wilson10,glaze15,velizcuba16}. Each observation $\theta_j$ contributes to the current estimate of $L_{n+1,\theta}$ via the probability $f_{\theta_j}(\theta)$ (Fig. \ref{fig2_preddist}B). Observations are weighted by assuming the observer has a fixed belief about the value $\ep$, specifying the average number of trials they expect the distribution $s_n(\theta)$ to remain the same. Leveraging techniques in probabilistic inference (See Methods), we find
\begin{linenomath}
\begin{align}
L_{n+1,\theta} = \bar{\PP}_0 \cdot \left[ \frac{(1- \ep)^n}{\PP(\theta_{1:n})} \prod_{j=1}^{n}  f_{\theta_j}(\theta) +   \ep \sum_{r=0}^{n-1}  \frac{(1- \ep)^r}{\PP(\theta_{n-r+1:n})} \prod_{j=n-r+1}^{n} f_{\theta_j}(\theta) \right],  \label{likesoln}
\end{align}
\end{linenomath}
where $\bar{\PP}_0: = 1/360$ is the uniform density for $-180^{\circ} \leq \theta< 180^{\circ}$. To understand Eq.~(\ref{likesoln}), it is instructive to examine limits of the parameter $\ep$ that admit approximations or exact updates. \\
\vspace{-3mm}

\noindent
{\bf Static environments ($\ep \to 0$).} In the limit $\ep \to 0$, the observer assumes the environment is static, so the predictive distribution is comprised of equal weightings of each observation (See Fig. \ref{fig2_preddist}C)~\cite{gold02,bogacz06}:
\begin{linenomath}
\begin{align}
L_{n+1,\theta} = \frac{\bar{\PP}_0}{\PP(\theta_{1:n})} \prod_{j=1}^n f_{\theta_j}(\theta). \label{static}
\end{align}
\end{linenomath}
As has been shown previously, Eq.~(\ref{static}) can be written iteratively~\cite{beck08}:
\begin{linenomath}
\begin{align*}
L_{n+1,\theta} = \frac{\PP(\theta_{1:n-1})}{\PP(\theta_{1:n})}f_{\theta_n}(\theta) L_{n,\theta},
\end{align*}
\end{linenomath}
suggesting such a computation could be implemented and represented by neural circuits. Temporal integration of tuned inputs has been demonstrated in both neural recordings~\cite{gold07,churchland08,kable09} and circuit models~\cite{machens05,bogacz06,beck08} of decision-making tasks. Most oculomotor delayed-response tasks use a distribution of targets $s(\theta)$ that is constant across trials~\cite{funahashi89,pesaran02,wimmer14,papadimitriou15}. Therefore, Eq.~(\ref{static}) is the optimal strategy for obtaining an estimate of $s(\theta)$, assuming the observer has a correct representation of the probability $f_{\theta_j}(\theta)$. For instance, if the distribution $s(\theta)$ were peaked, repeated observations $\theta_{1:n}$ would gradually improve the observer's estimate of that peak in Eq.~(\ref{static}). In changing environments ($\ep > 0$), recently observed targets are weighted more strongly than older targets, and the predictive distribution should down-weight the influence of past targets at a rate that increases with $\ep$~\cite{velizcuba16}. \\
\vspace{-3mm}


\noindent
{\bf Rapidly-changing environment ($\ep \approx 1$).} Our work focuses on the limit where the environment changes rapidly, $\ep \approx 1$ ($0<(1- \ep) \ll 1$), to account for biases that depend on the previous trial's target $\theta_n$ (See Methods for other cases). In this case, the predictive distribution for trial $n+1$ is a single peaked function centered at $\theta_n$ (Fig. \ref{fig2_preddist}C).
The observer assumes the environment changes fast enough that each subsequent target is likely drawn from a new distribution ($s_{n+1} (\theta) \not\equiv s_{n} (\theta)$). This is a suboptimal strategy, but matches the typical trends of interference in working memory. Applying this assumption to Eq.~(\ref{likesoln}), the formula for $L_{n+1,\theta}$ is dominated by terms of order $(1- \ep)$ and larger. Truncating to ${\mc O}(1-\ep)$ and normalizing the update equation (See Methods) then yields
\begin{linenomath}
\begin{align}
\tilde{L}_{n+1,\theta} = \ep \bar{\PP}_0 + (1-\ep) f_{\theta_n}(\theta). \label{approxn3}
\end{align}
\end{linenomath}
Thus, the dominant contribution from $\theta_{1:n}$ to $\tilde{L}_{n+1,\theta}$
is the target $\theta_n$ observed during the previous trial $n$ (Fig. \ref{fig2_preddist}C), similar to the behavioral findings of Papadimitriou et al. (2015)~\cite{papadimitriou15}.

Note, sequential computations are trivial in the limit of a constantly-changing environment $\ep \to 1$, since the observer assumes the environment is reset after each trial. Prior observations provide no information about the present distribution, so the predictive distribution is always uniform: $L_{n+1,\theta}  \equiv \bar{\PP}_0$.

In summary, a probabilistic inference model that assumes the distribution of targets is predictable over short timescales leads to response biases that depend mostly on the previous trial. We now demonstrate that this predictive distribution can be incorporated into a low-dimensional attractor model which describes the degradation of target memory during the delay-period of visual working memory tasks~\cite{brody03,renart03,kilpatrick13b}.

\subsection*{Incorporating suboptimal predictions into working memory}

We model the loading, storage, and recall of a target angle $\theta$ using a low-dimensional attractor model spanning the space of possible target angles $\theta \in [-180,180)^{\circ}$. These dynamics can be implemented in recurrent neuronal networks with slow excitation and broad inhibition~\cite{amari77,compte00,wimmer14}. Before examining the effects of neural architecture, we discuss how to incorporate the predictive distribution update, Eq.~(\ref{approxn3}), into an associated low-dimensional model. Our analysis links the update of the predictive distribution to the spatial organization of attractors in a network. Importantly, working memory is degraded by dynamic fluctuations, so the stored target angle wanders diffusively during the delay-period~\cite{compte00,kilpatrick13b,wimmer14}.

\begin{figure}[t]
\vspace{-12mm}
\begin{center} \includegraphics[width=13.2cm]{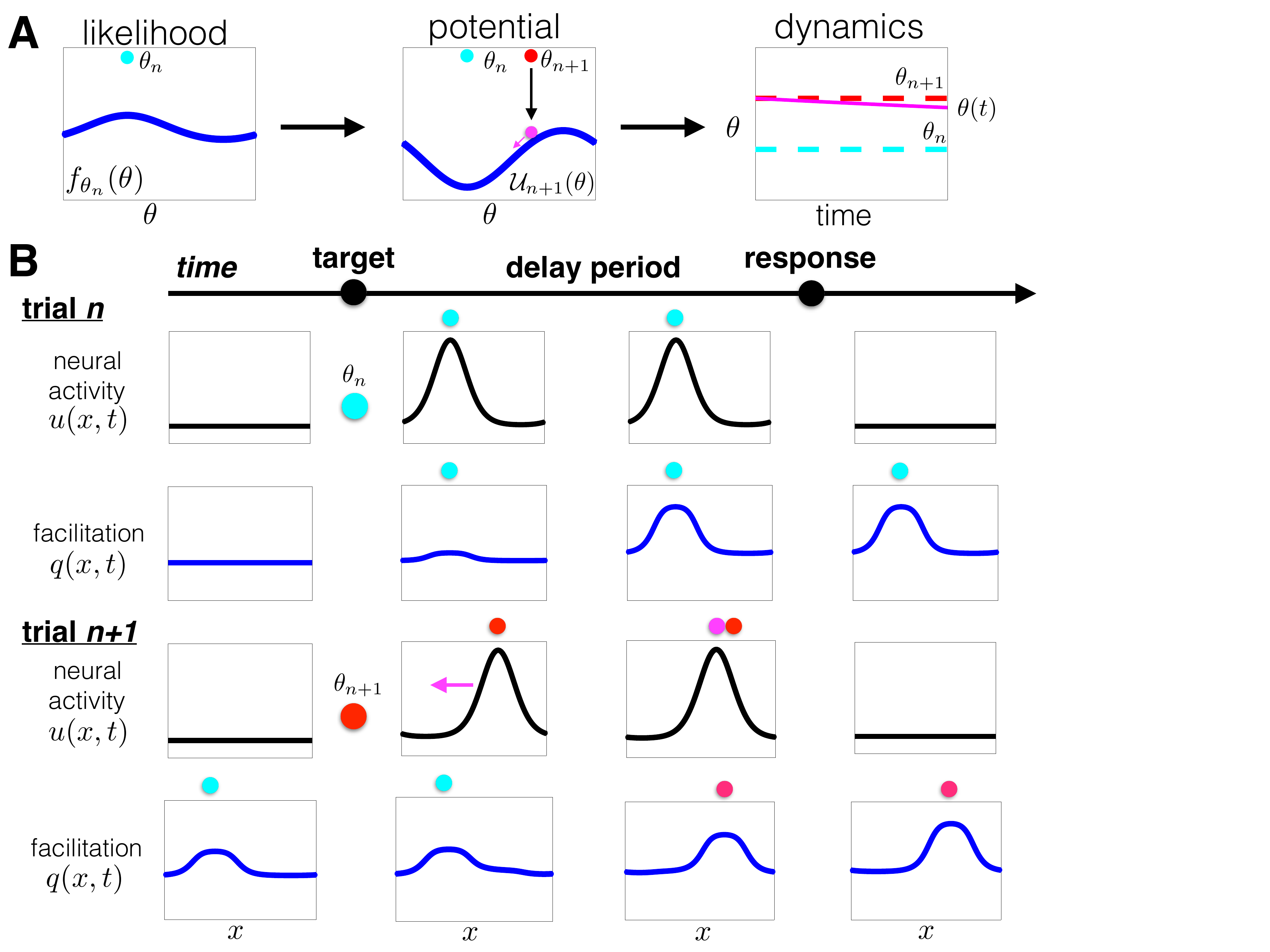} \end{center}
\caption{Encoding the predictive distribution in the potential function of an attractor network. (A) In a rapidly-changing environment, the predictive distribution is determined by the probability $f_{\theta_n}(\theta)$ (See Fig. \ref{fig2_preddist}C). In the low-dimensional system, with dynamics described by Eq.~(\ref{memsde}), this probability is represented by a potential function ${\mc U}_{n+1}(\theta)$ whose peak (valley) corresponds to the valley (peak) of $f_{\theta_n}(\theta)$, so the stored angle $\theta (t)$ drifts towards the minimum of ${\mc U}_{n+1}(\theta)$ during the delay-period. (B) A recurrent network with neurons distributed across $x \in [-180,180)^{\circ}$ with STF (Fig. \ref{fig1_task}E) can implement these dynamics. The position of the trial $n$ target is encoded by the spatial profile of STF $q(x,t)$ during the early portion of trial $n+1$, attracting the neural activity $u(x,t)$ bump during the delay-period.}
\label{fig3_potential}
\end{figure}



During the delay-period of a single trial, the stored target angle $\theta (t)$ evolves according to a stochastic differential equation~\cite{renart03}: 
\begin{linenomath}
\begin{align}
\d \theta (t) = - \frac{\d {\mc U}( \theta(t))}{\d \theta} \d t + \sigma_{\theta} \d \xi(t). \label{memsde}
\end{align}
\end{linenomath}
Here $\theta (t)$ is restricted to the periodic domain $\theta \in [-180,180)^{\circ}$ and $\d \xi$ is a standard white noise process. Eq.~(\ref{memsde}) can be derived as the low-dimensional projection for the location of a bump attractor in a recurrent network. The potential gradient $-{\mc U}'(\theta)$ models spatial heterogeneity in neural architecture that shapes attractor dynamics (Fig. \ref{fig3_potential}A). During trial $n+1$, we label the potential ${\mc U}(\theta) \equiv {\mc U}_{n+1}(\theta)$. Classic models of bump attractors on a ring assume distance-dependent connectivity~\cite{amari77,compte00}. The case ${\mc U}_{n+1}'(\theta) \not\equiv 0$ accounts for spatial heterogeneity in connectivity that may arise from a combination of training and synaptic plasticity~\cite{renart03}, or random components of synaptic architecture~\cite{wang06}.
The potential landscape can be updated during each trial, so at the beginning of trial $n+1$ it has the form ${\mc U}_{n+1}(\theta)$. When ${\mc U}_{n+1}( \theta) \equiv 0$, the potential is flat, so $\theta(t)$ evolves along a line attractor~\cite{brody03}. On the other hand, when the potential is heterogeneous, ${\mc U}_{n+1}(\theta) \not \equiv 0$, $\theta (t)$ tends to drift toward one of a finite number of discrete attractors~\cite{renart03,kilpatrick13b}. We will incorporate a process whereby previous targets are used to update the potential, so ${\mc U}_{n+1}(\theta)$ is typically heterogeneous.
The observer sees the target at the beginning of trial $n+1$, $\theta (0) = \theta_{n+1}$ (Fig. \ref{fig3_potential}A), and the angle $\theta (t)$ evolves according to Eq.~(\ref{memsde}) during the delay-period, lasting $T_D$ time units. After the delay-period, $\theta (T_D)$ is the recalled angle.
Depending on the underlying potential ${\mc U}_n(\theta)$, there will be a strong bias to a subset of possible targets.

We derive a correspondence between the probabilistic inference model and attractor model by assuming stationarity of ${\mc U}_{n+1}(\theta)$ within each trial (See Methods). In the recurrent network model (Fig. \ref{fig1_task}E), we take these within-trial dynamics into account. Freezing ${\mc U}_{n+1}(\theta)$ during a trial allows us to relate the statistics of the position $\theta (t)$ to the shape of the potential. Specifically, we relate the stationary density of Eq.~(\ref{memsde}) to the desired predictive distribution $L_{n+1,\theta}$ (See Methods). Thus, if information about the current trial's target $\theta_{n+1}$ is degraded, the probability distribution associated with the recalled target angle $\theta$ is $L_{n+1,\theta}$. Focusing on interference trends in Fig. \ref{fig1_task}, we aim to have the attractor structure of Eq.~(\ref{memsde}) represent the predictive distribution in Eq.~(\ref{approxn3}). Our calculations relate the potential function in trial $n+1$ to the probability generated by the trial $n$ target (Fig. \ref{fig3_potential}A) as
\begin{linenomath}
\begin{align}
{\mc U}_{n+1}(\theta) \propto - f_{\theta_n}(\theta).  \label{simplepot}
\end{align}
\end{linenomath}
The potential ${\mc U}_{n+1}(\theta)$ can be implemented by a decaying plasticity process that facilitates synapses from neurons tuned to the previous target $\theta_n$. As we will show, this can be accomplished via STF (Fig. \ref{fig3_potential}B).

\subsection*{Short-term facilitation generates interference in working memory}

We now show a neuronal network model describing neural activity $u(x,t)$ subject to STF $q(x,t)$ can incorporate predictive distribution updates derived above. Predictions are stored in the dynamically changing synaptic weights of a recurrent neuronal network.
The recurrent network model spatially labels neurons according to
their target orientation preference, determining the distance-dependent structure of inputs to the network. This is captured by a network with local excitation and effective inhibition that is fast and broad. Connectivity is shaped dynamically by STF (Fig. \ref{fig1_task}E).
See Methods for more details.


A sequence of delayed-response protocols is implemented in the recurrent network by specifying a spatiotemporal input $I(x,t)$ across trials (top of Fig. \ref{fig1_task}F). Each trial has a cue period of time length $T_C$; a delay-period of time length $T_D^{n+1}$; and an intertrial period of time length $T_I^{n+1}$ before the next target is presented. The network receives a peaked current centered at the neurons preferring the presented target angle $\theta_{n+1}$ during the cue period of trial $n+1$; no external input during the delay-period; and a strong inactivating current after the delay-period~\cite{compte00,wimmer14}.
The resulting bump attractor drifts in the direction of the bump from trial $n$, due to the STF at the location of the trial $n$ bump (Figs. \ref{fig1_task}F and \ref{fig3_potential}B).

\begin{figure}[t]
\begin{center} 
\includegraphics[width=12cm]{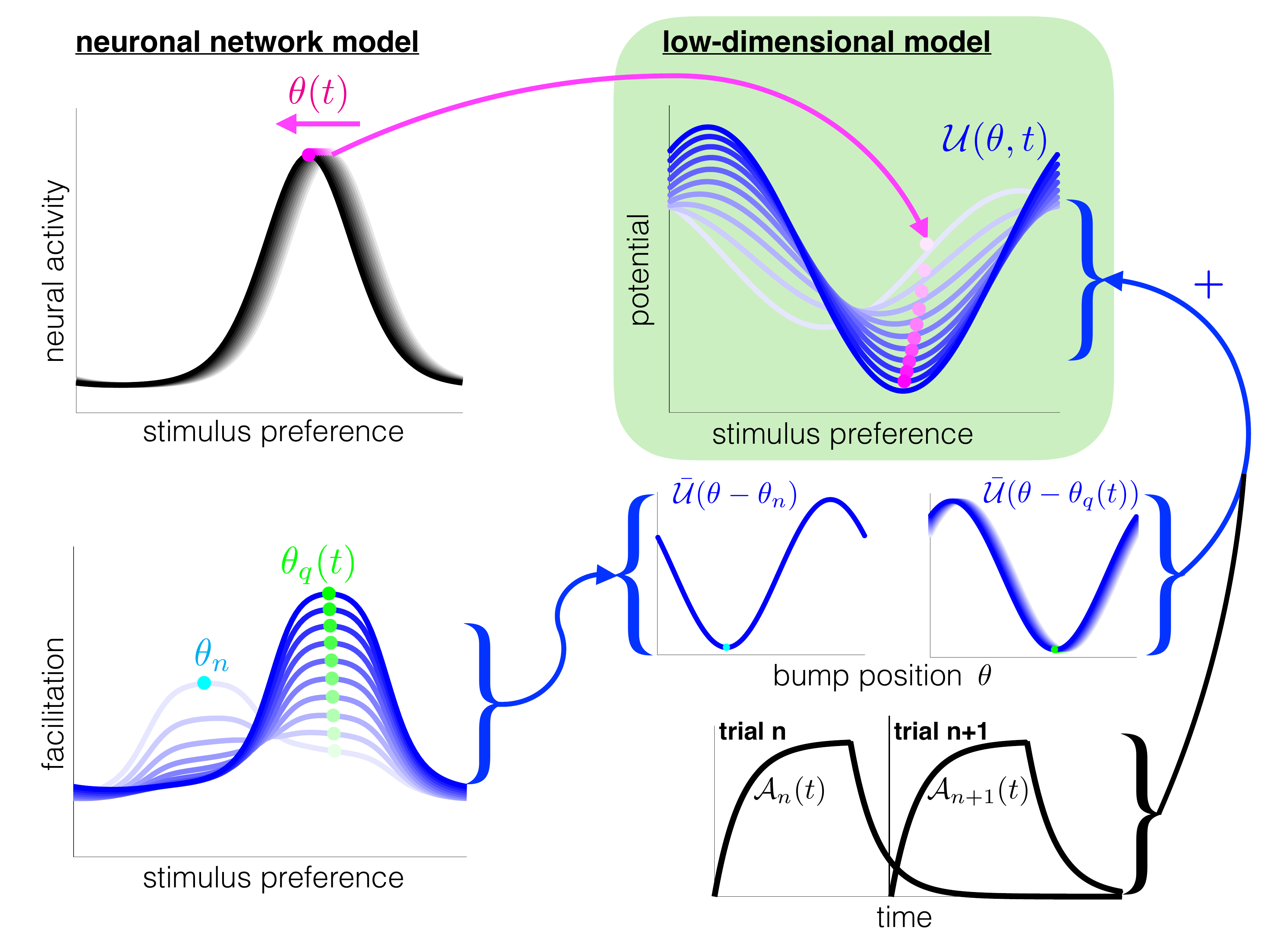}
\end{center}
\caption{Low-dimensional system (green box) captures the motion of the bump ($\theta (t)$) and the evolving potential, ${\mc U}(\theta,t)$, shaped by STF. The center-of-mass of the neural activity bump $\theta (t)$ is attracted by the most facilitated region of the network, ${\rm argmin}_{\theta} \left[ {\mc U}(\theta,t) \right]$. The current trial's bump location $\theta (t)$ attracts the variable $\theta_q(t)$, indicating the location of STF in the current trial. The evolving potential ${\mc U}(\theta,t)$ is then comprised of the weighted sum of the potential arising from the previous target ${\mc U}(\theta - \theta_n)$ and the current bump ${\mc U}(\theta - \theta_q(t))$. Dynamic fluctuations also perturb the position $\theta (t)$, so memory would degrade diffusively in the case of a flat potential ${\mc U}(\theta) \equiv 0$.  See Methods for a complete derivation.}
\label{fig4_lowdim}
\end{figure}

The mechanism underlying intertrial bias is determined by projecting our recurrent network model to a low-dimensional system that extends Eq~(\ref{memsde}) to account for STF. To reduce the recurrent network, we project the fast dynamics of bump solutions to an equation for the bump's position $\theta(t)$ in trial $n+1$~\cite{itskov11,kilpatrick13b}. The STF variable $q(x,t)$ determines an evolving potential function ${\mc U}(\theta,t)$ that shapes the bump's position (Fig. \ref{fig4_lowdim}). We use timescale separation methods (See Methods) to derive a set of stochastic differential equations, which approximates the motion of the bump's position $\theta (t)$ and the location of STF $\theta_q(t)$:
\begin{linenomath}
\begin{subequations} \label{lowdsys}
\begin{align}
\d \theta (t) &= - {\mc A}_n(t) \frac{\d \bar{\mc U}(\theta(t) - \theta_n)}{\d \theta} \d t -  {\mc A}_{n+1}(t)\frac{\d \bar{\mc U}(\theta(t) - \theta_q(t))}{\d \theta} \d t + \sigma \d \xi (t), \\
\tau \dot{\theta}_q (t) &= - d(\theta_q(t) - \theta (t)),
\end{align}
\end{subequations}
\end{linenomath}
during trial $n+1$ ($t_n < t< t_{n+1}$). The slowly-evolving potential gradient $- \frac{\pd}{\pd \theta} {\mc U}(\theta,t)$ shaping the dynamics of $\theta(t)$ is a mixture of STF contributions from trial $n$ (decaying ${\mc A}_n(t)$) and trial $n+1$ (increasing ${\mc A}_{n+1}(t)$). The bump position $\theta (t)$ moves towards the minimum of this dynamic potential, ${\rm argmin}_{\theta}\left[ {\mc U}(\theta,t) \right]$ (Fig. \ref{fig4_lowdim}). The variable $\theta_q(t)$ is the location of STF originating in trial $n+1$, and its position slowly moves toward the bump location $\theta (t)$. The parametrized timescale $\tau$ of STF is inversely related to the observer's perceived environmental change rate $\ep$ in Eq.~(\ref{approxn3}), since increasing $\ep$ corresponds to decreasing $\tau$.

The presence of STF provides two contributions to the slow dynamics of the bump position $\theta (t)$. The memory of the previous trial's target $\theta_n$ is reflected by the potential $\bar{\mc U}(\theta - \theta_n)$, whose effect decays slowly during trial $n+1$. This attracts $\theta (t)$, but the movement of $\theta (t)$ towards $\theta_n$ is slowed by the onset of the STF variable initially centered at $\theta_{n+1}$. The STF variable's center-of-mass $\theta_q(t)$ slowly drifts towards $\theta_n$, which allows $\theta (t)$ to drift there as well, $\bar{\mc U}(\theta - \theta_q(t))$. This accounts for the slow build-up of the bias that increases with the length of the delay-period~\cite{papadimitriou15}.

\subsection*{Target- and time-dependent trends match experimental observations}

We now demonstrate that the bias observed in the visual working memory experiments of Papadimitriou et al. (2015) can be accounted for by our recurrent network model (Fig. \ref{fig1_task}E) and our low-dimensional description of bump motion dynamics (Fig. \ref{fig4_lowdim}). To represent a sequence of working memory trials, targets $(\theta_1, \theta_2, \theta_3,...)$ were presented to the recurrent network, and the center-of-mass of the bump was recorded at the end of each delay-period, representing the response $(r_1, r_2, r_3,...)$ (See Methods). The bias of responses was determined by computing the difference between the response and the presented target, $r_n - \theta_n$. Means and variances of the bias were determined under each condition.

\begin{figure}[t]
\vspace{-2mm}
\begin{center} \includegraphics[width=15cm]{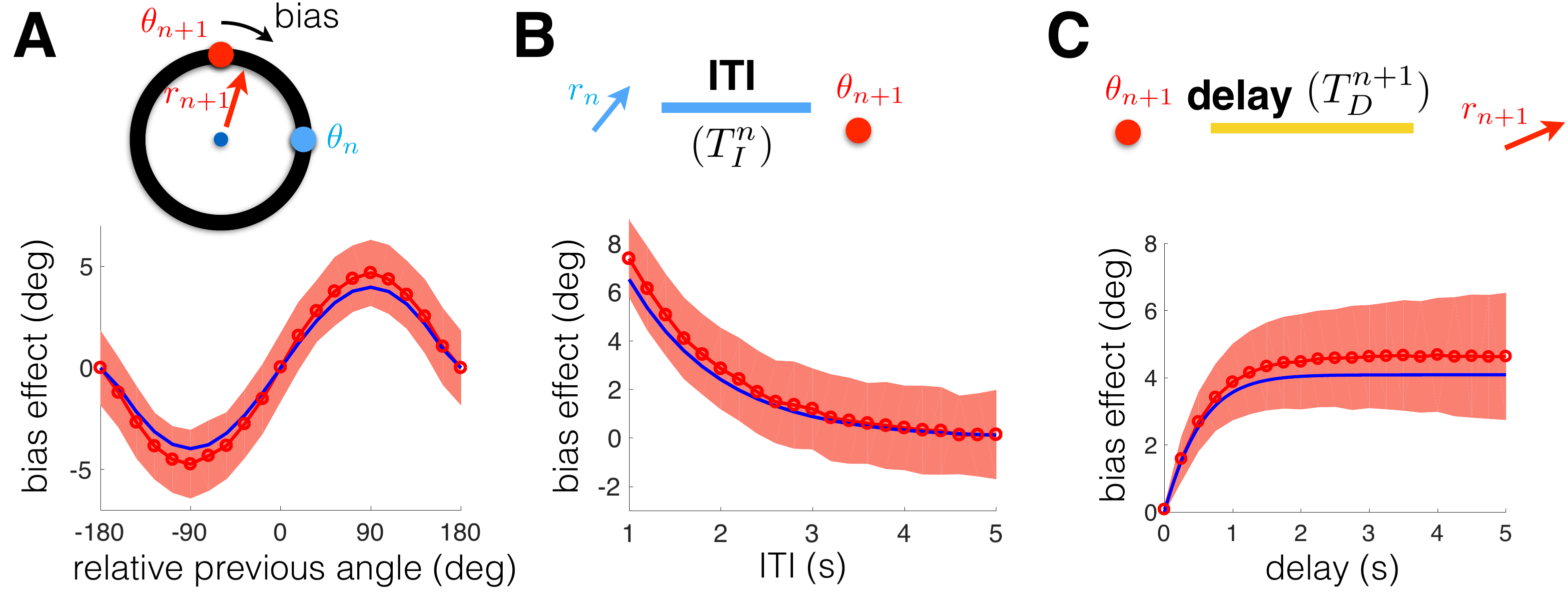}  \end{center}
\caption{Intertrial bias is shaped by (A) the angle between targets $\theta_{n+1}$ and $\theta_n$; (B) the interval between trials $n$ and $n+1$ (ITI); and (C) the delay-period during trial $n+1$. (A) Responses in trial $n+1$ are biased in the direction of the previous trial target ($\theta_n$).
Simulations of the recurrent network (red circles) are compared with the low-dimensional model (blue line). Shaded region indicates one standard deviation (See Methods for details). (B) The peak bias decreases with the intertrial interval (ITI), due to the temporal decay of STF. (C) The peak bias increases with the delay since the bump drifts towards the equilibrium position determined by the STF profile.}
\label{fig5_stats}
\end{figure}

Our results are summarized in Fig. \ref{fig5_stats}, focusing on three conditions considered by Papadimitriou et al. (2015)~\cite{papadimitriou15}. First, we calculated the bias when conditioning on the angle between the trial $n$ and trial $n+1$ targets, $\theta_n - \theta_{n+1}$ (Fig. \ref{fig5_stats}A). Positive (negative) angles lead to positive (negative) bias; i.e. the bump drifts in the direction of the previous target $\theta_n$. The bias is graded with the difference, $\theta_n - \theta_{n+1}$. To expose this effect, we averaged across trials, since the recurrent network incorporates dynamic input fluctuations, as in bump attractor models of visuospatial working memory~\cite{compte00,wimmer14}. We also calculated the peak bias as a function of the intertrial interval (ITI), the time between the trial $n$ response ($r_n$) and the trial $n+1$ target presentation $\theta_{n+1}$. Consistent with Papadimitriou et al. (2015)~\cite{papadimitriou15}, the peak bias decreased with the ITI (Fig. \ref{fig5_stats}B). The mechanism for this decrease is the slow decay in the STF of synapses utilized in the previous trial. Finally, the peak bias increased with the delay within a trial, since persistent activity was slowly attracted to the location of the previous target (Fig. \ref{fig5_stats}C). This slow saturation arises due to the slow kinetics of STF within a trial.

Not only did our recurrent network model recapitulate the main findings of Papadimitriou et al. (2015)~\cite{papadimitriou15}, we also found our low-dimensional description of the bump and STF variable dynamics had these properties (blue curves in Fig. \ref{fig5_stats}). The mechanics underlying the bias can be described with a model of a particle evolving in a slowly changing potential (Fig. \ref{fig4_lowdim}), shaped by the dynamics of STF. Having established a mechanism for the bias, we consider how different protocols determine the statistics of responses, not conditioned with sequential trial information.

\subsection*{Task protocol shapes ensemble statistics}

\begin{figure}[t]
\vspace{-2mm}
\begin{center}  \includegraphics[width=15cm]{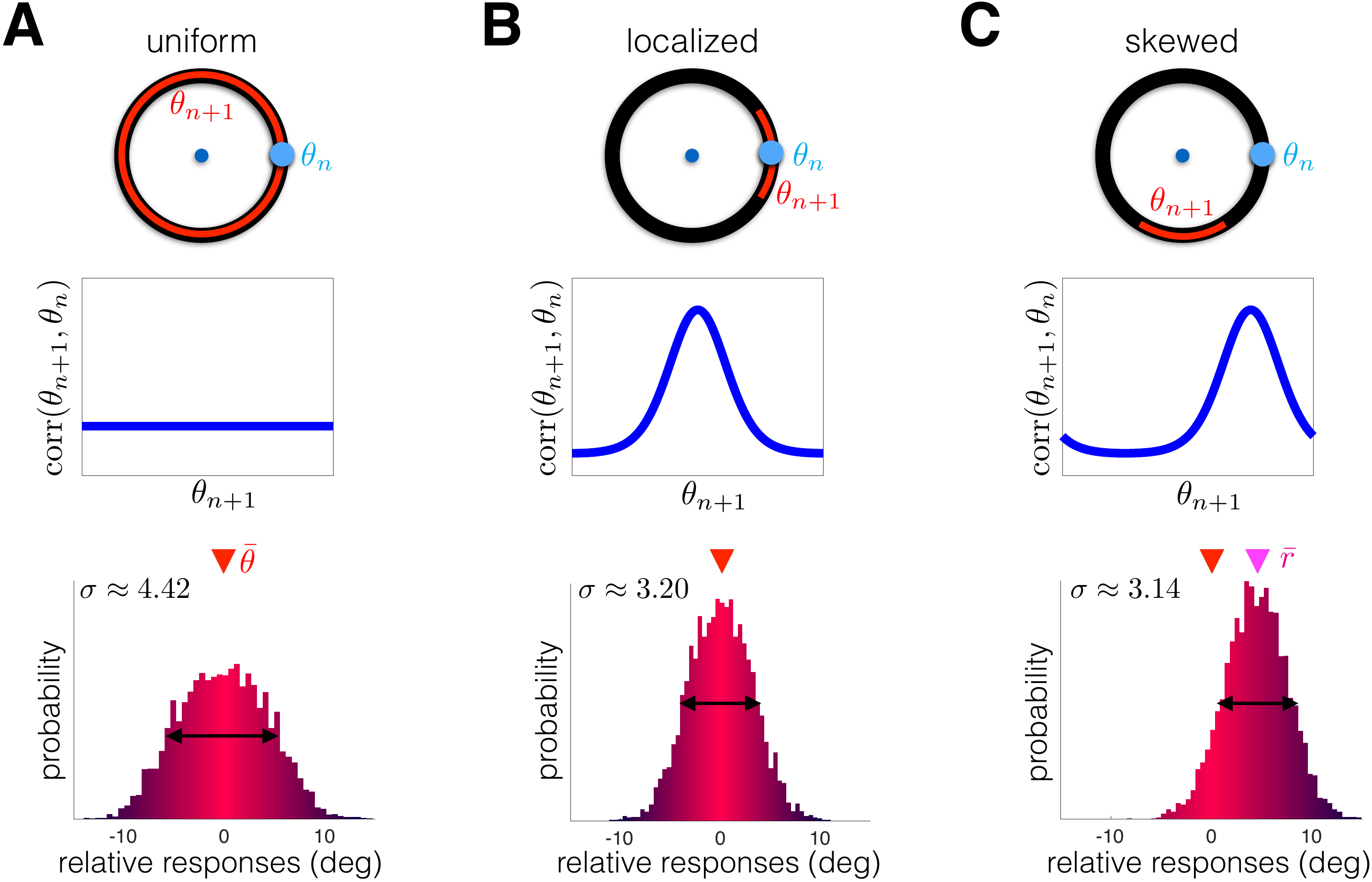}  \end{center}
\caption{Response distribution is shaped by correlations between target angles in adjacent trials ${\rm corr}(\theta_{n+1}, \theta_n)$. (A) Visual working memory protocols typically use a sequence of targets with no trial-to-trial correlations (uniform ${\rm corr} (\theta_{n+1},\theta_{n})$, shown for $\theta_n \equiv 0^{\circ}$)~\cite{compte00,wimmer14}. Relative responses ($r_n - \theta_n$) are normally distributed about the true target. (B) Current target $\theta_{n+1}$ is correlated with the previous target $\theta_{n}$ according to a locally peaked distribution. The response distribution narrows (note decreased standard deviation $\sigma$), since the target $\theta_{n+1}$ is often close to the previous target $\theta_n$. (C) Current target $\theta_{n+1}$ is skewed clockwise from previous angle $\theta_{n}$. Responses are thus skewed counter-clockwise towards the previous target (note average response $\bar{r}$ is shifted). Numerical methods are described in Methods.}
\label{fig6_ensemble}
\end{figure}

Visual working memory tasks are often designed such that sequential target locations are uncorrelated~\cite{compte00,wimmer14}. In such protocols, there is no advantage in using previous trial information to predict targets within the current trial. Nonetheless, these biases persist in the intertrial response correlations discussed in Papadimitriou et al. (2015)~\cite{papadimitriou15} and Fig. \ref{fig5_stats}. On the other hand, interference might be advantageous for tasks with correlations between successive target angles, $\theta_n$ and $\theta_{n+1}$. Consider object motion tracking tasks with transiently occluded objects~\cite{bennett06}, so the object's location prior to occlusion predicts its subsequent location when it reappears. Memory of object location that persists beyond a single trial can therefore be useful for naturally-inspired tasks.

We demonstrate this idea by comparing the network's performance in working memory tasks whose targets are drawn from distributions with different intertrial correlation profiles (Fig. \ref{fig6_ensemble}). As a control, we consider the case of uniform correlations between target $\theta_n$ and target $\theta_{n+1}$ (Fig. \ref{fig6_ensemble}A). Responses are normally distributed about the true target angle. The dynamics of the bump encoding the target are shaped by both input fluctuations and a bias in the direction of the previous target on individual trials. However, the directional bias is not apparent in the entire distribution of response angles, since it samples from all possible pairs $(\theta_{n+1}, \theta_n)$. An ensemble-wide measure of performance is given by the standard deviation of the response distribution ($\sigma \approx 4.42$). When target angles are correlated between trials, the relative response distribution narrows (Fig. \ref{fig6_ensemble}B). Memory of the previous trial's target $\theta_n$ stabilizes the memory of the current trial's target $\theta_{n+1}$, decreasing the standard deviation of responses ($\sigma \approx 3.20$). There is a high probability the current target $\theta_{n+1}$ will be close to the previous target $\theta_n$, so the timescale of the network's underlying inference process is reasonably well matched to the environment. However, such effects can be deleterious when the previous angle $\theta_{n}$ is skewed in comparison to the current angle $\theta_{n+1}$. Protocols with this correlation structure lead to a systematic bias in the relative response distribution, so its peak is shifted away from zero (Fig. \ref{fig6_ensemble}C).

Our neuronal network model predicts that, if an intertrial bias is present in the computations of a neural circuit, it should be detectable by varying the intertrial correlation structure of target angles $\theta_n$. Furthermore, when there are strong local correlations between adjacent trials (${\rm corr}( \theta_{n+1}, \theta_{n})$ is large for $|\theta_{n+1} - \theta_{n}|$ small), responses are more accurate than for protocols with uncorrelated adjacent trial angles. Since the strength of the bias increases as the intertrial interval is decreased, due to the decay of STF, we expect the effect to be more pronounced for trials taken closer together.

\subsection*{Two timescales of memory degradation}


\begin{figure}[t]
\vspace{-2mm}
\begin{center} \includegraphics[width=14cm]{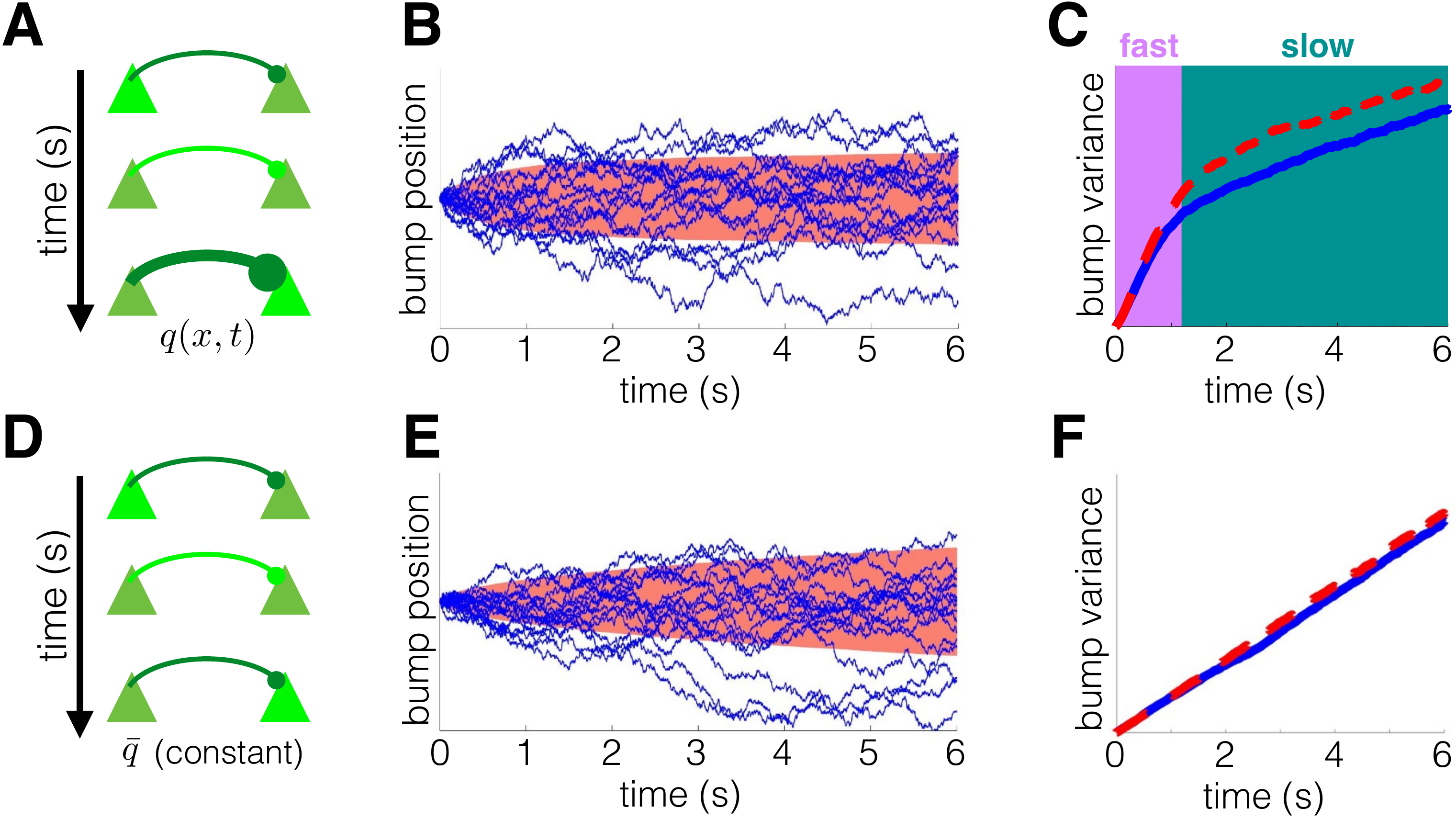}   \end{center}
\caption{Recurrent networks with STF (panels A-C) exhibit two timescales of delay-period dynamics, in contrast to the single timescale dynamics of networks with static synapses (panels D-F). (A) Release of neurotransmitter leads to the strengthening of the synapse via STF. (B) In a facilitating network, bump trajectories (lines) stray less from their initial position due to the attractive effect of STF. Large ensemble standard deviation shown in red. (C) STF generates two phases of variance scaling. An initial fast phase is followed by a slower phase due to the dampening effect of STF in both neuronal network (red dashed) and low-dimensional (blue solid) simulations. (D) Network with static synapses. (E) Bump trajectories obey linear diffusion, due to the spatial homogeneity of the network. (F) Variance grows linearly with time, a hallmark of pure diffusion. See Methods for further details.}
\label{fig7_compare}
\end{figure}

Wimmer et al. (2014) have shown that both the normal distribution of saccade endpoints and observed changes in neural firing rates during the delay-period can be replicated by a diffusing bump attractor model~\cite{wimmer14}. We have shown that a recurrent network with STF (Fig. \ref{fig1_task}E) still leads to a normal distribution of predicted response angles (Fig. \ref{fig6_ensemble}A). Our model also provides new predictions for the dynamics of memory degradation, which we now compare with the standard diffusing bump attractor model~\cite{compte00} (Fig. \ref{fig7_compare}). In a network with STF (Fig. \ref{fig7_compare}A), bump trajectories evolve in a history-dependent fashion (Fig. \ref{fig7_compare}B). Initially, bumps diffuse freely, but are eventually drawn to their starting location by facilitated synapses (See also Fig. \ref{fig4_lowdim}). This leads to two distinct phases of diffusion, as shown in plots of the bump variance (Fig. \ref{fig7_compare}C). Rapid diffusion occurs initially, as the bump equilibrates to the quasistationary density determined by the slowly evolving potential (Fig. \ref{fig4_lowdim}). Slower diffusion occurs subsequently, as spatial heterogeneity in synaptic architecture gradually responds to changes in bump position via STF. Stabilizing effects of STF on bump attractors have been analyzed previously~\cite{itskov11}, but our identification of these multiple timescale dynamics is novel. This feature of the bump dynamics is not present in networks with static synapses (Fig. \ref{fig7_compare}D). Here, bumps evolve as a noise-driven particle over a flat potential landscape (Fig. \ref{fig7_compare}E), described by Brownian motion: a memoryless stochastic process~\cite{brody03}. Variance in bump position scales purely linearly with time (Fig. \ref{fig7_compare}F), and the diffusion coefficient can be computed using a low-dimensional approximation~\cite{kilpatrick13b}.

The qualitative differences between the bump attractor with and without dynamic synapses should be detectable in both behavioral and neurophysiological recordings~\cite{wimmer14}. Moreover, the observed intertrial bias identified in recent analyses of behavioral data requires some mechanism for transferring information between trials that is distinct from neural activity~\cite{papadimitriou15}, as dynamic synapses are in our model. In total, our model provides both an intuition for the behavioral motivation as well as neurophysiological mechanisms that produce such interference.

\section*{Discussion}


Neural circuit models of visual working memory tend to use neural activity variables as the encoders of target locations. Our computational models account for interference in visual working memory using both suboptimal Bayesian inference and STF acting on a recurrent network model of delay-period activity. The timescale and prior target dependence of attractive biases we observe correspond to psychophysical observations of behavioral experiments in monkeys~\cite{papadimitriou15}. STF evolves dynamically over seconds~\cite{wang06}, matching the kinetics of interference in visual working memory. The link we have drawn between our two models suggests neural circuits can implement probabilistic inference using short-term plasticity.

\subsection*{Experimental predictions}

More complete descriptions of the neural mechanics of visual working memory could be accomplished by analyzing the effects of correlations in sequential target presentations. Since responses in subsequent trials are shaped by the previous trial's target~\cite{papadimitriou15}, computational models can be validated by determining how well their response distributions reflect trial-to-trial target correlations (Fig. \ref{fig6_ensemble}). It is also possible that introducing target sequences whose distributions change in time could impact quantitative features of interference. For instance, implementing tasks with target sequences that have multiple trial correlations may extend the timescale of interference beyond a single trial. Furthermore, our model predicts that multiple timescales emerge in the statistics of delay-period activity during a working memory task (Fig. \ref{fig7_compare}). Variance of recall error increases sublinearly in our model, consistent with a recent reanalysis of psychophysical data of saccades to remembered visual targets~\cite{white94,qi15}. The dynamics of our model are thus inconsistent with the purely linear diffusion of recall error common in bump attractor models with static synapses~\cite{compte00,wimmer14}.

The idea that STF may play a role in working memory is not new~\cite{barak07,mongillo08}, and there is evidence that prefrontal cortex neurons exhibit dynamic patterns of activity during the delay-period, suggestive of an underlying modulatory process~\cite{stokes13}. However, it remains unclear how the presence of STF may shape the encoding of working memories. Our model suggests STF can transfer attractive biases between trials. Recent findings on the biophysics of STF could be harnessed to examine how blocking STF shapes behavioral biases in monkey experiments~\cite{jackman16}. We predict that reducing the effects of the STF would both decrease the systematic bias in responses and increase the amplitude of errors, since the stabilizing effect of STF on the persistent activity will be diminished~\cite{itskov11}.

\subsection*{Alternative neurophysiological mechanisms for intertrial bias}

Our study accounts for biases observed by Papadimitriou et al. (2015)~\cite{papadimitriou15}, who identified an attraction between the previous target and current response. Strengthening synapses that originate from recently active neurons can attract neural activity states in subsequent trials. This is consistent with recent experiments showing latent and ``activity-silent" working memories can be reactivated in humans using transcranial magnetic stimulation~\cite{rose16}, suggesting working memory is maintained by mechanisms other than target-tuned persistent neural activity~\cite{mongillo08,stokes13}. The principle of using short-term plasticity to store memories of visual working memory targets could be extended to account for longer timescales and more intricate statistical structures. Short-term depression (STD) could effect a repulsive bias on subsequent responses, since neural activity would be less likely to persist in recently-activated depressed regions of the network. In this way, STD could encode a predictive distribution for targets that are anti-correlated to the previously present target.

Other physiological mechanisms could also shape network responses to encode past observations in a predictive distribution. Long-term plasticity is a more viable candidate for encoding predictive distributions that accumulate observations over long timescales. Consider a protocol that uses the same distribution of target angles throughout an entire experiment, but this distribution is biased towards a discrete set of possible angles~\cite{kilpatrick13b}. For a recurrent network to represent this distribution, it must retain information about past target presentations over a long timescale. Many biophysical processes underlying plasticity are slow enough to encode information from such lengthy sequences~\cite{benna16}. Furthermore, the distributed nature of working memory suggests that there may be brain regions whose task-relevant neural activity partially persists from one trial to the next~\cite{constantinidis16}. Such activity could shape low-level sensory interpretations of targets in subsequent trials.

\subsection*{Synaptic plasticity can stabilize working memory}

Several modeling studies of working memory have focused on the computational capability of synaptic dynamics~\cite{barak14}.
For instance, STF can prolong the lifetime of working memories in spatially heterogeneous networks, since facilitated synapses slow the systematic drift of bump attractor states~\cite{itskov11}. This is related to our finding that STF reduces the diffusion of bumps in response to dynamic fluctuations (Fig. \ref{fig7_compare}B), generating two timescales of memory degradation, corresponding to the bump variance (Fig. \ref{fig7_compare}C). This scaling may be detectable in neural recordings or behavioral data, since recall errors may saturate if stabilized by STF. Facilitation can also account for experimentally observed increases in spike train irregularity during the delay-period in circuit models that support tuned persistent activity~\cite{hansel13}. Alternatively, homeostatic synaptic scaling can compensate for spatial heterogeneity, which would otherwise cause persistent states to drift~\cite{renart03}. However, the short homeostatic timescales often suggested in models do not often match experimental observations~\cite{zenke17}.

Models of working memory have also replaced persistent neural firing with stimulus-selective STF, so neuronal spiking is only required for recall at the end of the delay-period~\cite{mongillo08}. One advantage of this model is that multiple items can be stored in the dynamic efficacy of synapses, and the item capacity can be regulated by external excitation for different task load demands~\cite{mi17}. Our model proposes that STF plays a supporting rather than a primary role, and there is extensive neurophysiological evidence corroborating persistent neural activity as a primary working memory encoder~\cite{wimmer14}.

\subsection*{Robust working memory via excitatory/inhibitory balance}

Computational modeling studies have demonstrated that a balance of fast inhibition and slow excitation can stabilize networks, so they accurately integrate inputs~\cite{machens05}. Drift in the representation of a continuous parameter can be reduced by incorporating negative-derivative feedback into the firing rate dynamics of a network, similar to introducing strong friction into the mechanics of particle motion on a sloped landscape~\cite{lim13}. Fast inhibition balanced by slower excitation produces negative feedback that is proportional to the time-derivative of population activity. A related mechanism can be implemented in spiking networks wherein fast inhibition rapidly prevents runaway excitation, and the resulting network still elicits highly irregular activity characteristic of cortical population discharges~\cite{boerlin13}. Mutually inhibiting balanced networks are similarly capable of representing working memory of continuous parameters~\cite{shaham17}, and extending our framework by incorporating STF into this paradigm would be a fruitful direction of future study.

\subsection*{Extensions to multi-item working memory}

Working memory can store multiple items at once, and the neural mechanisms of interference between simultaneously stored items are the focus of ongoing work~\cite{ma14}. While there is consensus that working memory is a limited resource allocated across stored items, controversy remains over whether resource allocation is quantized (e.g., slots)~\cite{luck13} or continuous (e.g., fluid)~\cite{ma14}.  Spatially-organized neural circuit models can recapitulate inter-item biases observed in multi-item working memory experiments, and provide a theory for how network interactions produce such errors~\cite{wei12}. In these models, each remembered item corresponds to an activity bump, and the spatial scale of lateral inhibition determines the relationship between recall error and item number~\cite{bays15}. The model provides a theory for attractive bias and forgetting of items since nearby activity bumps merge with one another. This is related to the mechanism of attractive bias in our model, but a significant difference is that previous models only required localized excitation whereas we use STF. It would be interesting identify the temporal dynamics of biases in multi-item working memory to see if they require slower timescale processes like short-term plasticity.

\subsection*{Tuning short-term plasticity to the environmental timescale}

We have not identified a mechanism whereby our network model's timescale of inference could be tuned to learn the inherent timescale of the environment. There is recent evidence from decision-making experiments that humans can learn the timescale on which their environment changes and use this information to weight their observations toward a decision~\cite{glaze15}. Our model suggests that the trial-history inference utilized by subjects in Papadimitriou et al. (2015) is significantly suboptimal~\cite{papadimitriou15}, perhaps because it is difficult to infer the timescale of relevant past-trial information. The complexity, sensitivity, and resource expense of optimal inference in most contexts likely makes it impossible to implement exactly in neural circuits~\cite{austerweil15}. This may explain why humans often use suboptimal methods for accumulating evidence~\cite{beck12}. Plasticity processes that determine the timescale of evidence accumulation may be shaped across generations by evolution, or across a lifetime of development. Nonetheless, metaplasticity processes can internally tune the dynamics of plasticity responses in networks without changing synaptic efficacy itself, and these changes could occur in a reward-dependent way~\cite{hulme14}. Recently, a model of reward-based metaplasticity was proposed to account for adaptive learning observed in a probabilistic reversal learning task~\cite{farashahi17}. Such a process could modify the timescale and other features of short-term plasticity in ways that improve task performance in working memory as well.

\subsection*{Conclusions}

Our results suggest that interference observed in visual working memory tasks can be accounted for by a persistently active neural circuit with STF.
Importantly, interference is graded by the time between trials and during a trial. The interplay of synaptic and neural processes involved in interference may have arisen as a robust system for processing visual information that changes on the timescale of seconds. More work is need to determine how information about the environment stretches across multiple timescales to shape responses in cognitive tasks. We expect that identifying the neural origin of such biases will improve our understanding of how working memory fits into the brain's information-processing hierarchy.

\section*{Online Methods}

%
%
%

\subsection*{Assumptions of the inference model}

Our model performs nonparametric density estimation to approximate the distribution $s_{n+1}(\theta)$ from which a target $\theta$ will be drawn before trial $n+1$. The observer assumes the possible distributions $s(\theta)$ are drawn from a function space $s \in S$ according to the prior $p(s)$. We assume that marginalizing over all such distributions yields the uniform density $\bar{\PP}_0 = \int_{S} s (\theta) p(s) \d s  = 1/360$. One possibility is that the distribution $s_{n+1}(\theta)$ is constructed by drawing $N$-tuples ${\bf a}$ and $\boldsymbol{\psi}$ from a uniform distribution over the hypercubes $[0,a_{max}]^N$ and $[-180^{\circ},180^{\circ})^N$ and using the entries to construct an exponential distribution of a sum of cosines:
\begin{linenomath}
\begin{align*}
s_{n+1}(\theta ) = {\mc N}_{s} \exp \left[ \sum_{j=1}^{N} a_j \cos (\omega_j \cdot (\theta - \psi_j)) \right],
\end{align*}
\end{linenomath}
where $\omega_j = j \pi/180$ and ${\mc N}_s$ is a normalization constant. For instance, when $N=1$,
\begin{linenomath}
\begin{align*}
s_{n+1}(\theta ) = {\mc N}_{s} \exp \left[a_1 \cos (\omega_1 \cdot (\theta - \psi_1)) \right],
\end{align*}
\end{linenomath}
peaked at $\psi_1$. For the main instantiation and reduction of our model, knowing the specific family of distributions is unnecessary.

\begin{table}
\begin{tabular}{l|l}
\multicolumn{2}{c}{} \\
\hline 
symbol & description \\
\hline
$\theta_n$ & target observed on trial $n$\\
$s_{n+1}(\theta)$ & unknown distribution of targets observer is attempting to infer before trial $n+1$ \\
$f_{\theta_n}(\theta)$ & probability of observing target $\theta$ in the present trial \\
$\ep$ & assumed change rate of the environment observer uses to discount evidence\\
$L_{n+1,\theta}$ & predictive distribution: observer's expected probability of seeing target $\theta$ in trial $n+1$,  \\
& given their past observations $\theta_{1:n}$ and an assumed change rate $\ep$ of the environment \\
$r_n$ & run-length: number of trials preceding trial $n+1$ with the same target distribution \\
$\bar{\PP}_0$ & uniform prior: distribution for which all targets $\theta \in [-180^{\circ}, 180^{\circ})$ have same probability \\
\hline 
\end{tabular}
\caption{Variables and parameters of the probabilistic inference model. \label{table1}}
\end{table}

The probability $f_{\theta'}(\theta) := \PP (\theta_{n+1} = \theta | \theta_n = \theta' , s_{n+1}(\theta) \equiv s_n(\theta))$ is defined under static conditions ($s_{n+1}(\theta) \equiv s_n(\theta)$) to separate the dynamic effects of sampling distribution $s_n(\theta)$ changes. We are performing nonparametric Bayesian estimation of the distribution, and the probability $f_{\theta'}(\theta)$ is already marginalized over the space of distributions $s(\theta)$. Thus, we do not model the intermediate step of inferring the probability of each distribution $s (\theta)$ and marginalizing, but it could be computed by integrating over the prior on the function space, $f_{\theta'}(\theta) = \int_S s(\theta) f(s| \theta') \d s$. Each observation $\theta'$ would give the probability $f(s|\theta')$ that the current distribution is $s(\theta)$. Integrating over the space of all distributions $s \in S$ provides the probability the next target will be $\theta$, based on the previous observation $\theta'$ alone and the assumption that the distribution remains the same from trial $n$ to $n+1$. Details on the difference between parametric and nonparametric Bayesian estimation of densities can be found in Orbanz and Teh (2011)~\cite{orbanz11}. Note, we assume self-conjugacy of $f_{\theta'}(\theta) = f_{\theta}(\theta')$, which follows since the order of observations does not matter while the environment remains fixed. This relationship will also make the predictiveness of our model more apparent. It is important to note that the observer assumes the form of $f_{\theta'}(\theta)$, but this is not necessarily the distribution an ideal observer should use.
For illustration, we consider a family of distributions given by an exponential of cosines:
\begin{linenomath}
\begin{align}
f_{\theta'}(\theta) = {\mc N}_{\theta} \exp \left[ \sum_{j=1}^{N} a_j \cos (\omega_j \cdot ( \theta - \theta')) \right], \label{expcos}
\end{align}
\end{linenomath}
which is self-conjugate: $f_{\theta'}(\theta) \equiv f_{\theta}(\theta')$~\cite{diaconis79}. A distribution like Eq.~(\ref{expcos}) would emerge from a generative model with distance-dependent spatial correlations in the ensemble of produced targets. The example $f_{\theta'}(\theta)$ we use for comparison with our recurrent network with STF is close to the case of Eq.~(\ref{expcos}) with $N=1$. A description of the parameters and variables in our model is provided in Table \ref{table1}.

\subsection*{Derivation of the probabilistic inference model}

The observer's predictive distribution $L_{n+1,\theta} = \PP( \theta_{n+1}| \theta_{1:n}, \ep)$ is derived by computing the probability of observing $\theta_{n+1}$ given each prior observation $\theta_j$, $j=1,...,n$. Importantly, we must compute the probability of each run length $r_n = r$, $r=0,...,n$, corresponding to the number of trials the assumed underlying distribution $s_n(\theta)$ has remained the same~\cite{adams07,wilson10}. Knowing the probability of each run length will inform us of how much to weight each observation $\theta_j$, $j=1,...,n$. In particular, $r_n = n$ indicates the environment has remained the same since the first trial, and $r_n =0$ indicates the environment changes between trial $n$ and $n+1$. Summing over all possible run lengths, the marginal predictive distribution is
\begin{linenomath}
\begin{align}
L_{n+1,\theta} = \sum_{r=0}^{n} \PP (\theta_{n+1}| r_n=r , \theta_{1:n}^r) \PP(r_n=r|\theta_{1:n}, \ep), \label{recupdate1}
\end{align}
\end{linenomath}
where $\PP(\theta_{n+1}|r_n=r,  \theta_{1:n}^r)$ is the conditional predictive distribution assuming run length $r_n = r$ and $\PP(r_n=r|\theta_{1:n})$ is the conditional probability of the run length $r_n=r$ given the series of target angles $\theta_{1:n}$. We further simplify Eq.~(\ref{recupdate1}) as follows: First, utilizing sequential analysis, we find that if the present run length is $r_n=r$, the conditional predictive distribution is given by the product of probabilties from the last $r$ observations~\cite{velizcuba16}:
\begin{align}
\PP( \theta_{n+1}| r_n = r, \theta_{1:n}^r) =  \frac{\bar{\PP}_0}{\PP(\theta_{n-r+1:n})} \prod_{j=n-r+1}^{n}  f_{\theta_j}(\theta),  \label{margpred}
\end{align}
where again $\bar{\PP}_0$ is the uniform distribution and we have utilized our self-conjugacy assumption for $f_{\theta'}(\theta) \equiv f_{\theta}(\theta')$. Next, we assume that observations provide no information about the present run length $r_n$, which would be a consequence of the observer making no a priori assumptions on the overall distribution from which targets $\theta_{1:n}$ are drawn. Thus, the observer only uses their knowledge of the change rate of the environment $\ep$ to determine the probability of a given run length $r_n=r$, and the conditional probability can be computed
\begin{align}
\PP (r_n=r | \theta_{1:n}, \ep) = \PP(r_n = r | \ep) = \left\{ \begin{array}{cl} \ep (1- \ep)^r, & r < n, \\   (1- \ep)^n, & r=n. \end{array} \right. \label{runform}
\end{align}
Plugging Eqs.~(\ref{margpred}--\ref{runform}) into the update Eq.~(\ref{recupdate1}), we find the probability of the next target being at angle $\theta_{n+1}=\theta$, given that the previous $n$ targets were $\theta_{1:n}$, is:
\begin{linenomath}
\begin{align*}
L_{n+1,\theta} = \bar{\PP}_0 \cdot \left[ \frac{(1- \ep)^n}{\PP(\theta_{1:n})} \prod_{j=1}^{n}  f_{\theta_j}(\theta) +   \ep \sum_{r=0}^{n-1}  \frac{(1- \ep)^r}{\PP(\theta_{n-r+1:n})} \prod_{j=n-r+1}^{n} f_{\theta_j}(\theta) \right]. 
\end{align*}
\end{linenomath}

\subsection*{Limit of slowly-changing environment (small $\ep$)}

Here, we examine the case $0< \ep \ll1 $, where the environment changes very slowly. Assuming independence of the target angles selected on each trial $\theta_{1:n}$~\cite{bogacz06},  $\PP (\theta_{n-r:n}) = \PP(\theta_{n-r:n-1}) \PP(\theta_n)$, we can split the probabilities over the target sequences $\theta_{n-r:n}$ into products: $\PP( \theta_{n-r:n}) = \prod_{j=n-r}^{n} \PP(\theta_j) = \bar{\PP}_0^{r+1}$. The last equality holds since the family of possible distributions $s_n(\theta )$ averages to a constant $\bar{\PP}_0$, the uniform density. Applying this assumption to Eq.~(\ref{likesoln}) and truncating to ${\mc O}(\ep)$, we have
\begin{linenomath}
\begin{align*}
\tilde{L}_{n+1,\theta} = {\mc N}_s \cdot \left[ (1 - n \ep) \prod_{j=1}^{n} \frac{f_{\theta_j}(\theta)}{\bar{\PP}_0} + \ep \sum_{r=0}^{n-1} \prod_{j=n-r+1}^n \frac{f_{\theta_j}(\theta)}{\bar{\PP}_0} \right],
\end{align*}
\end{linenomath}
noting we must choose ${\mc N}_s$ so $\int_{-180}^{180} L_{n+1,\theta} \d \theta =1$, normalized at each step.

\subsection*{Limit of rapidly-changing environment ($\ep \approx 1$)}

Here, we examine the case $\ep \approx 1$ ($0 < (1-\ep) \ll 1$), a rapidly-changing environment. Applying this assumption to Eq.~(\ref{likesoln}), we find $L_{n+1,\theta}$ is dominated by terms of order $(1- \ep)$ and larger. Terms of order $(1-\ep)^2$ are much smaller. For instance, we can approximate to linear order,
dropping terms of ${\mc O}((1- \ep)^2)$,
to reduce Eq.~(\ref{likesoln}) to
\begin{linenomath} 
\begin{align}
L_{n+1,\theta} \approx \ep \bar{\PP}_0 \left[ 1 + \frac{1-\ep}{\PP(\theta_n)} f_{\theta_n}(\theta) \right]. \label{approxn}
\end{align}
\end{linenomath}
Furthermore,
we ensure the expression in Eq.~(\ref{approxn}) is normalized by writing
\begin{linenomath}
\begin{align*}
\tilde{L}_{n+1,\theta} = \frac{\bar{\PP}_0 + (1-\ep) f_{\theta_n}(\theta)}{2 - \ep},
\end{align*}
\end{linenomath}
since $ \int_{-180}^{180}\left[  \bar{\PP}_0 +  (1-\ep) f_{\theta_n}(\theta) \right] \d \theta = 2-\ep$. Alternatively, we can truncate by multiplying through by $[1- (1-\ep)]/[1 - (1-\ep)]$, truncating to ${\mc O}(1-\ep)$ and normalizing to yield
\begin{linenomath}
\begin{align*}
\tilde{L}_{n+1,\theta} = \ep \bar{\PP}_0 + (1-\ep) f_{\theta_n}(\theta),
\end{align*}
\end{linenomath}
the key update equation in our Results (Figs. \ref{fig2_preddist} and \ref{fig3_potential}A). Higher order approximations are obtained by keeping more terms from Eq.~(\ref{likesoln}); e.g., a second order approximation yields
\begin{linenomath}
\begin{align*}
L_{n+1,\theta} &\approx \ep \bar{\PP}_0 + \ep (1-\ep) f_{\theta_n}(\theta) + \frac{\ep (1- \ep)^2 \bar{\PP}_0}{\PP(\theta_{n-1:n})} f_{\theta_n}(\theta) f_{\theta_{n-1}}(\theta),
\end{align*}
\end{linenomath}
successively downweighting the influence of previous observations ($\theta_{n-1}$). 

\subsection*{Relating the predictive distribution to the potential of an attractor model}

\begin{table}
\begin{tabular}{l|l}
\multicolumn{2}{c}{}  \\
\hline
symbol & description  \\ \hline
$\theta (t)$ & observer's estimate of the remembered target during the delay-period \\
${\mc U}_{n+1}(\theta)$ & potential determining the systematic dynamics of $\theta$ during the delay-period \\
$\sigma_{\theta}$ & standard deviation of the dynamic fluctuations \\
$p_{n+1}(\theta,t)$ & probability density function (pdf) for finding the observer's estimate at $\theta$ at time $t$ \\
$\bar{p}_{n+1}(\theta)$ & stationary density: pdf of observer's estimate $\theta$ in the long time limit, $t \to \infty$ \\
\hline 
\end{tabular}
\caption{Variables and parameters of the particle evolving on a potential model. \label{table2}}
\end{table}

A predictive distribution can be represented by an attractor model by first determining the formula of the stationary distribution of Eq.~(\ref{memsde}), given an arbitrary potential function ${\mc U}_{n+1}(\theta)$. Eq.~(\ref{memsde}) can be reformulated as an equivalent Fokker-Planck equation for the represented angle $\theta$ during trial $n+1$ assuming the present potential function is ${\mc U}_{n+1}(\theta)$~\cite{risken96},
\begin{linenomath}
\begin{align}
\frac{\pd p_{n+1} ( \theta, t)}{\pd t} = \frac{\pd}{\pd \theta} \left[ \frac{\d {\mc U}_{n+1}( \theta)}{\d \theta} p_{n+1}(\theta, t) \right] + \frac{\sigma_{\theta}^2}{2} \frac{\pd^2 p_{n+1}(\theta,t)}{\pd \theta^2}, \label{memfp}
\end{align}
\end{linenomath}
where $p_{n+1}(\theta, t)$ is the probability density corresponding to the target angle estimate $\theta$ at time $t$. The initial estimate of the target is exact, $\theta (0) = \theta_{n+1}$, so $p_{n+1}(\theta,0) = \delta (\theta - \theta_{n+1})$ is the initial condition. We summarize the constituent variables and model parameters in Table \ref{table2}.

We now derive the form of ${\mc U}_{n+1}(\theta)$ that leads to a stationary density corresponding the predictive distribution $L_{n+1,\theta}$ in the limit $t \to \infty$ in Eq.~(\ref{memfp}). The stationary density $\bar{p}_{n+1}(\theta)$ is analogous to a predictive distribution represented by Eq.~(\ref{memsde}) since it is the probability the system represents when no information about the current trial's target $\theta_{n+1}$ remains.
Thus, we build a rule to update ${\mc U}_{n+1}(\theta)$ to mirror the update of $L_{n+1,\theta}$ in Eq.~(\ref{approxn3}). To obtain this result, we match the stationary density for Eq.~(\ref{memfp}) to the updated predictive distribution:
\begin{linenomath}
\begin{align}
\lim_{t \to \infty} p_{n+1}(\theta, t) = \bar{p}_{n+1}(\theta) = L_{n+1,\theta}.  \label{stattopred}
\end{align}
\end{linenomath}
Solving Eq.~(\ref{memfp}) for its stationary solution, we find that during trial $n+1$:
\begin{linenomath}
\begin{align}
\bar{p}_{n+1}(\theta) = \chi_{n+1} \exp \left[ - \frac{2 {\mc U}_{n+1}(\theta)}{\sigma_{\theta}^2} \right], \label{statdist}
\end{align}
\end{linenomath}
where $\chi_{n+1}$ is a normalization factor chosen so that $\int_{-180}^{180} \bar{p}_{n+1}(\theta) \d \theta = 1$. Plugging Eq.~(\ref{statdist}) into Eq.~(\ref{stattopred}) and solving for ${\mc U}_{n+1}(\theta)$, we obtain
\begin{linenomath}
\begin{align*}
{\mc U}_{n+1}(\theta) = \frac{\sigma_{\theta}^2}{2} \ln \frac{\chi_{n+1}}{L_{n+1,\theta}}.
\end{align*}
\end{linenomath}
For a rapidly changing environment $0<(1- \ep) \ll 1$, we approximate $L_{n+1,\theta}$ using Eq.~(\ref{approxn3}) so that
\begin{linenomath}
\begin{align*}
{\mc U}_{n+1}(\theta) &= \frac{\sigma_{\theta}^2}{2} \left[ \ln \chi_{n+1} - \ln \left( \ep \bar{\PP}_0 + (1 - \ep) f_{\theta_n} ( \theta) \right) \right] \\
& \approx \frac{\sigma_{\theta}^2}{2} \left[ \ln \frac{\chi_{n+1}}{\bar{\PP}_0} - (1- \ep) \frac{f_{\theta_n}(\theta) - \bar{\PP}_0}{\bar{\PP}_0} \right],
\end{align*}
\end{linenomath}
where we have linearized in $(1- \ep)$. However, for Eq.~(\ref{memsde}), only the derivative of ${\mc U}_{n+1}(\theta)$ impacts the dynamics, so we drop the additive constants and examine the proportionality
\begin{linenomath}
\begin{align*}
{\mc U}_{n+1}(\theta) \propto - f_{\theta_n}(\theta).
\end{align*}
\end{linenomath}
In the limit of weak interactions between trials, the potential ${\mc U}_{n+1}(\theta)$ should be shaped like the negative of the probability $f_{\theta_n}(\theta)$ based on the previous trial's target $\theta_n$.

\subsection*{Bump attractor model with short-term facilitation}

\begin{table}
\begin{tabular}{l|l}
\multicolumn{2}{c}{}  \\
\hline
symbol & description  \\ \hline
$u(x,t)$ & synaptic input to location $x$ at time $t$ \\
$q(x,t)$ & short term facilitation (STF) variable at location $x$ and time $t$: increases the strength of \\
&  connectivity  from neurons at location $x$ in response to their activation  \\
$\tau_u $ & timescale of synaptic excitation, set to 10ms in simulations~\cite{hausser97}  \\
$w(x-y)$ & baseline strength and polarity of effective connectivity from neurons at location $y$ to $x$  \\
$F(u)$ & nonlinearity that converts synaptic input to output firing rate of local neural population \\
$J(x,t)$ & spatiotemporal input: external input due to visual stimuli plus dynamic fluctuations \\
$\tau$ & timescale of STF, set to 1ms in simulations~\cite{tsodyks97} \\
$\beta$ & onset rate of STF, set to 0.01 in simulations~\cite{mongillo08} \\
$q_+$ & maximal increase in synaptic utilization, set to 2 in simulations~\cite{mongillo08}  \\
\hline 
\end{tabular}
\caption{Variables and parameters of the recurrent network model. \label{table3}}
\end{table}

Our neuronal network model is comprised of two variables evolving in space $x\in [-180,180)^{\circ}$, corresponding to the stimulus preference of neurons at that location, and time $t>0$. Variables and parameters are summarized in Table \ref{table3}, and the evolution equations consist of one stochastic integrodifferential equation and one auxiliary differential equation:
\begin{linenomath}
\begin{subequations}  \label{bumpnet}
\begin{align}
\tau_u \d u (x, t) & = \left[ - u(x, t) + \int_{- 180}^{180} w(x-y) (1+q(y,t)) F(u(y,t)) \d y + I(x,t) \right] \d t + \d W (x,t), \label{bnet1} \\
\tau \dot{q} (x, t) &= -q(x,t) + \beta F(u(x,t)) (q_+ - q(x,t)), \label{bnet2}
\end{align}
\end{subequations}
\end{linenomath}
where $u( x, t)$ describes the evolution of the normalized synaptic input at location $x$. The model Eq.~(\ref{bumpnet}) can be derived as the large system size limit of a population of synaptically coupled spiking neurons~\cite{bressloff12}, and similar dynamics have been validated in spiking networks with lateral inhibitory connectivity~
\cite{compte00,wimmer14}. We fix the timescale of dynamics by setting $\tau_u = 10$ms, so time evolves according to units of a typical excitatory synaptic time constant~\cite{hausser97}. This population rate model can be explicitly analyzed to link the architecture of the network to a low-dimensional description of the dynamics of a bump attractor as described by Eq.~(\ref{memsde}).

Each location $x$ in the network receives recurrent coupling defined by the weight function $w(x - y)$. We take this function to be peaked when $x=y$ and decreasing as the distance $|x-y|$ grows, in line with anatomical studies of delay-period neurons in prefrontal cortex~\cite{goldmanrakic95}. We do not separately model excitatory and inhibitory populations, but Eq.~(\ref{bumpnet}) can be derived from a model with distinct excitatory and inhibitory populations in the limit of fast inhibitory synapses~\cite{amari77,carroll14}. Thus, we have combined excitatory and inhibitory populations, so $w(x-y)$ takes on both positive and negative values. Our analysis can be applied to a general class of distance-dependent connectivity functions, given by an arbitrary sum of cosines $w(x-y) = \sum_{n=0}^{\infty} \alpha_n \cos ( \omega_n (x-y) )$ where $\omega_n = n \pi/180$, and we will use a single cosine to illustrate in examples: $w(x-y) = \cos (\omega_1 (x-y) )$. The nonlinearity $F(u)$ converts the normalized synaptic input $u(x,t)$ into a normalized firing rate, $F(u) \in [0,1]$. We take this to be sigmoidal $F(u) = 1/\left[ 1 + \e^{-\gamma(u-\kappa)} \right]$~\cite{wilson73}, with a gain of $\gamma = 20$ and a threshold of $\kappa = 0.1$ in numerical simulations. In the high-gain limit ($\gamma \to \infty$), a Heaviside step function $F(u) = H(u - \kappa)$ allows for explicit calculations~\cite{amari77,bressloff12}.

Recurrent coupling is shaped by STF in active regions of the network ($F(u)>0$), as described by the variable $q(x,t) \in [0,q_+]$; $q_+>0$ and $\beta$ determine the maximal increase in synaptic utilization and the rate at which facilitation occurs~\cite{tsodyks97,tsodyks98}. For our numerical simulations, we consider the parameter values $q_+ = 2$ and $\beta = 0.01$, consistent with previous models employing facilitation in working memory circuits~\cite{itskov11,mongillo08,mi17} and experimental findings for facilitation responses in prefrontal cortex~\cite{wang06}. The timescale of plasticity is slow, $\tau = 1000{\rm ms} \gg 10$ms, consistent with experimental measurements~\cite{tsodyks97}. Our qualitative results are robust to parameter changes. Information from the previous trial is maintained by the slow-decaying kinetics of the facilitation variable $q(x,t)$, even in the absence of neural activity~\cite{mongillo08,mi17}.

Effects of the target and the response are described by the deterministic spatiotemporal input $I(x,t)$, which we discuss more in detail below. The noise process $W(x,t)$ is white in time and has an increment with mean $\langle \d W (x, t) \rangle \equiv 0$ and spatial correlation function $\langle \d W (x, t) \d W (y,s) \rangle = C (x - y) \delta (t - s) \d t \d s$. In numerical simulations, we take our correlation function to be $C(x-y) = \sigma_W^2 \cos (x-y)$ with $\sigma_W = 0.005$, so the model recapitulates the typical 1-5\% standard deviation in saccade endpoints observed in oculomotor delayed-response tasks with delay-periods from 1-10s~\cite{funahashi89,white94,wimmer14}. 

\subsection*{Implementing sequential delayed-response task protocol}

A series of oculomotor delayed-response tasks is executed by the network Eq.~(\ref{bumpnet}) by specifying a schedule of peaked inputs occurring during the cue periods of length $T_C$, no input during trial $n$'s delay-period of length $T_D^n$, and brief and strong inhibitory input of length $T_A$ after the response has been recorded, and then no input until the next trial. This is described by the spatiotemporal function
\begin{linenomath}
\begin{align*}
I(x,t) = \left\{ \begin{array}{lll} I_0 \exp \left[ I_1 (\cos (x - \theta_n)-1) \right], & t \in [t_n,t_n+T_C), \\ 0 , & t \in [t_n+T_C, t_n+T_C+T_D^n),  \\ -I_R, & t \in [t_n+T_C+T_D^n,t_n+T_C+T_D^n+T_A), \\ 0, & t \in [t_n+T_C+T_D^n+T_A,t_{n+1}),  \end{array} \right.
\end{align*}
\end{linenomath}
for all $n=1,2,3,...$, where $t_n$ is the starting time of the $n^{\text{th}}$ trial which has cue period $T_C$, delay-period $T_D^n$, inactivation period $T_A$, and subsequent intertrial interval $T_I^n$. Note that the delay and intertrial interval times may vary trial-to-trial, but the cue is always presented for the same period of time as in Papadimitriou et al. (2015)~\cite{papadimitriou15}. The amplitude of the cue-related stimulus is controlled by $I_0$, and $I_1$ controls is sharpness. Activity from trial $n$ is ceased by the global inactivating stimulus of amplitude $I_R$.

In numerical simulations, we fix the parameters $T_C = 500$ms; $T_A = 500$ms; $I_0 = 1$; $I_1 = 1$; and $I_R = 2$. Target locations $\theta_n$ are drawn from a uniform probability mass function (pmf) for the discrete set of angles $\theta_n \in \{ -180^{\circ}, -162^{\circ}, ..., 162^{\circ}\}$ to generate statistics in Fig. \ref{fig5_stats}A, which adequately resolves the bias effect curves for comparison with the results in Papadimitriou et al. (2015)~\cite{papadimitriou15}. Intertrial intervals are varied to produce Fig. \ref{fig5_stats}B by drawing $T_I^n : = t_{n+1} -(T_C+T_D^n+T_A)$ randomly from a uniform pmf for the discrete set of times $T_I^n \in \{1000,1200,...,5000\}$ms and $\theta_n$ randomly as in Fig. \ref{fig5_stats}A and identifying the $\theta_n$ that produces the maximal bias for each value of $T_I^n$. Delay-periods are varied to produce Fig. \ref{fig5_stats}C by drawing $T_D^n$ randomly from a uniform pmf for the discrete set of times $T_I^n \in \{0,200,...,5000\}$ms and following a similar procedure to Fig. \ref{fig5_stats}B. Draws from a uniform density function $\PP (\theta_n) \equiv \bar{\PP}_0$, defined on $\theta_n \in [-180,180)^{\circ}$ are used to generate the distribution in Fig. \ref{fig6_ensemble}A and plots in Fig. \ref{fig7_compare}. Nontrivial correlation structure in target selection is defined by the sum of a von Mises distribution and uniform distribution ${\rm corr} (\theta_{n+1}, \theta_{n}) = (1- \ep) {\mc N}_v \e^{25 \cos (\theta_n - \theta_{n+1} - \mu)} + \ep \bar{\PP}_0$ for fixed $\theta_n$ with $\ep = 0.5$; $\mu = 0$ for local correlations (Fig. \ref{fig6_ensemble}B) and $\mu = 90$ for skewed correlations (Fig. \ref{fig6_ensemble}C). 

The recurrent network, Eq.~(\ref{bumpnet}), is assumed to encode the initial target $\theta_{n}$ during trial $n$ via the center-of-mass $\theta(t)$ of the corresponding bump attractor. Representation of the cue at the end of the trial is determined by performing a readout on the neural activity $u(x,t)$ at the end of the delay time for trial $n$: $t = t_n + T_C + T_D^n$. One way of doing this would be to compute a circular mean over $x$ weighted by $u(x,t)$, but since $u(x,t)$ is a roughly symmetric and peaked function in $x$, computing $\theta(t):={\rm argmax}_x u(x,t)$ (when $t \in [t_n,t_n+T_C+T_D^n)$) is an accurate and efficient approximation~\cite{kilpatrick13b,wimmer14}. The bias and relative saccade endpoint on each trial $n$ are then determined by computing the difference $\theta (t) - \theta_n$ (Figs. \ref{fig5_stats}, \ref{fig6_ensemble}, and \ref{fig7_compare}).

\subsection*{Deriving the low-dimensional description of bump motion}

We analyze the mechanisms by which STF shapes the bias on subsequent trials by deriving a low-dimensional description for the motion of the bump position $\theta (t)$. To begin, note that in the absence of facilitation ($\beta \equiv 0$), the variable $q(x,t) \equiv 0$. In the absence of noise ($W(x,t) \equiv 0$), the resulting deterministic Eq.~(\ref{bumpnet}) has stationary bump solutions that are well studied and defined by the implicit equation~\cite{amari77,bressloff12,kilpatrick13}:
\begin{linenomath}
\begin{align*}
U(x) = \int_{-180}^{180} w(x - y) F(U(y)) \d y.
\end{align*}
\end{linenomath}
Assuming the stimulus $I(x,t)$ presented during the cue period of trial $n$ ($t \in [t_n,t_n+T_C)$) is strong enough to form a stationary bump solution, the impact of the facilitation variable $q(x,t)$ and noise $W(x,t)$ on $u(x,t)$ during the delay-period ($t \in [t_n+T_C,t_n+T_C+T_D^n)$) can be determined perturbatively, assuming $|q| \ll 1$ and $|\d W| \ll 1$. Since $\tau \gg \tau_u$, $u(x,t)$ will rapidly equilibrate to a quasi-steady-state determined by the profile of $q(x,t)$. We thus approximate the neural activity dynamics as $u(x,t) \approx U(x - \theta (t)) + \Phi (x,t)$, where $\theta (t)$ describes the dynamics of the bump center-of-mass during the delay-period ($|\theta| \ll 1$ and $|\d \theta| \ll 1$), and $\Phi(x,t)$ describes perturbations to the bump's shape ($|\Phi| \ll 1$). Plugging this approximation into Eq.~(\ref{bumpnet}) and truncating to linear order yields
\begin{linenomath}
\begin{align}
\d \Phi(x,t)- {\mc L}\Phi(x,t) \d t &=  U'(x)\d \theta + \int_{-180}^{180} w(x - y) q(y + \theta,t_s) F(U(y)) \d y \d t + \d W, \label{pertphi} 
\end{align}
\end{linenomath}
where ${\mc L}u = -u + \int_{-180}^{180} w(x- y) F'(U(y)) u(y) \d y$ is a linear operator and $q(x,t_s)$ is the facilitation variable evolving on the slow timescale $t_s = \tau_u t/\tau \ll t$, quasi-stationary on the fast timescale of $u(x,t)$. We ensure a bounded solution by requiring the right hand side of Eq.~(\ref{pertphi}) is orthogonal to the nullspace $V(x)$ of the adjoint linear operator ${\mc L}^*v = -v + F'(U) \int_{-180}^{180} w(x - y) v(y) \d y$. Orthogonality is enforced by requiring the inner product $\langle u,v \rangle = \int_{-180}^{180} u(x) v(x) \d x$ of the nullspace $V(x)$ with the inhomogeneous portion of Eq.~(\ref{pertphi}) is zero. It can be shown $V(x) = F'(U(x))U'(x)$ spans the nullspace of ${\mc L}^*$~\cite{kilpatrick13}. This yields the following equation for the evolution of the bump position:
\begin{linenomath}
\begin{align}
\d \theta (t) =  K(\theta (t),t_s) \d t + \sigma \d \xi (t),  \label{bumpsde}
\end{align}
\end{linenomath}
where the slowly evolving nonlinearity
\begin{linenomath}
\begin{align}
K (\theta, t_s) = \frac{\int_{-180}^{180} \int_{-180}^{180} w(x - y) q(y + \theta,t_s) F(U(y)) \d y F'(U(x)) U'(x) \d x}{\int_{-180}^{180} U'(x)^2 F'(U(x)) \d x}  \label{Kform}
\end{align}
\end{linenomath}
is shaped by the form of $q(x,t_s)$ and the noise $\xi (t)$ is a standard Wiener process that comes from filtering the full spatiotemporal noise process $\d W(x,t)$, so the diffusion coefficient
\begin{linenomath}
\begin{align*}
D: = \frac{\sigma^2}{2} = \frac{\int_{-180}^{180} \int_{-180}^{180} V(x) C(x-y) \d y \d x}{\left[ \int_{-180}^{180} U'(x) V(x) \d x \right]^2}.
\end{align*}
\end{linenomath}

Eq.~(\ref{bumpsde}) has the same form as Eq.~(\ref{memsde}).
Thus, if the facilitation variable $q(x,t_s)$ evolves trial-to-trial such that $K(\theta, t_s)$ has similar shape to $\displaystyle - \frac{\d {\mc U}_{n+1}}{\d \theta} (\theta)$ at the beginning of the $(n+1)^{\text{th}}$ trial ($t = t_{n+1}$), the dynamics of the network Eq.~(\ref{bumpnet}) can reflect a prior distribution based on the previous target(s). Given the approximation we derived in Eq.~(\ref{simplepot}), we enforce proportionality $K(\theta,t_{n+1}) \propto - \frac{\d {\mc U}_{n+1}}{\d \theta} (\theta)$:
\begin{linenomath}
\begin{align}
K(\theta,t_{n+1}) = \alpha \frac{\d f_{\theta_n}(\theta)}{\d \theta},  \label{Ktof}
\end{align}
\end{linenomath}
where $\alpha$ is a scaling constant and $t_{n+1}$ is the starting time of trial $n+1$ in the original time units $t = \tau t_s/ \tau_u$. The form of the probability $f_{\theta'}(\theta)$ that can be represented is therefore restricted by the dynamics of the facilitation variable $q(x,t)$. We can perform a direct calculation to identify how $q(x,t)$ relates to the predictive distribution it represents in the following special case.

\subsection*{Explicit solutions for high-gain firing rate nonlinearities}

To explicitly calculate solutions, we take the limit of high-gain, so that $F(u) \to H(u - \kappa)$ and $w(x) = \cos (\omega_1 x)$, note $\omega_1 = 180/\pi$. In this case, the bump solution $U(x-x_0) = (2 \sin (a)/ \omega_1) \cos (\omega_1(x-x_0))$ for $U(\pm a) = \kappa$ and null vector $V(x-x_0) = \delta (x-x_0-a) - \delta (x-x_0+a)$ (without loss of generality we take $x_0 \equiv 0$)~\cite{kilpatrick13}. Furthermore, we can determine the form of the evolution of $q(x,t)$ by studying the stationary solutions to Eq.~(\ref{bumpnet}) in the absence of noise ($W \equiv 0$). For a bump $U(x)$ centered at $x_0=0$, the associated stationary form for $Q(x)$ assuming $H(U(x)-\kappa)=1$ for $x \in (-a,a)$
and zero otherwise is $Q(x) = \beta q_+/(1+\beta)$ for $x \in (-a,a)$ and zero otherwise. Thus, if the previous target was at $\theta_n$, we expect $q(x,t)$ to have a shape resembling $Q(x-\theta_n)$ after trial $n$.  Assuming the cue plus delay time during trial $n$ was $T_C + T_D^n$ and the intertrial interval is $T_I^n$, slow dynamics will reshape the amplitude of $q(x,t)$ so ${\mc A}_n(T^n) = (1- \e^{-(T_C+T_D^n)/\tau}) \e^{-T_I^n/\tau}$ ($T^n=T_C+T_D^n+T_I^n$ is the total time block of each trial) and so $q(x,t) \approx {\mc A}_n(T^n) \cdot Q(x-\theta_n)$ at the beginning of trial $n+1$. A lengthy calculation of Eq.~(\ref{Kform}) combined with the relation Eq.~(\ref{Ktof}) yields:
\begin{linenomath}
\begin{align*}
\alpha \frac{\d f_{\theta_n}(\theta)}{\d \theta} = \frac{\beta q_+ {\mc A}_n(T^n)}{2 (1+ \beta) \tan(a) } \left[ {\rm sign} (\theta - \theta_n)(1- \cos (\omega_1(\theta- \theta_n))) - \tan(a) \sin (\omega_1(\theta- \theta_n)) \right],
\end{align*}
\end{linenomath}
for $|\theta - \theta_n| < 2a$,  and $\frac{\d f_{\theta_n}(\theta)}{\d \theta} \equiv 0$ otherwise. Integrating, we find this implies
\begin{linenomath}
\begin{align*}
f_{\theta_n}(\theta) \propto |\theta - \theta_n| - \sin | \theta - \theta_n| + \tan (a) \cos ( \theta - \theta_n), 
\end{align*}
\end{linenomath}
for $|\theta - \theta_n| < 2a$, and $f_{\theta_n}(\theta)$ constant otherwise. Thus, the STF dynamics allows the network architecture to represent a predictive distribution that is peaked at the previous target location (Fig. \ref{fig3_potential}). The amplitude of the $\theta$-dependent portion of the predictive distribution during trial $n+1$ is then controlled by cue, delay, and intertrial times ($T_C, T_D^{n+1}, T_I^{n+1}$) and the facilitation parameters ($\beta$, $q_+$, $\tau$). 

To derive a coupled pair of equations (Fig. \ref{fig4_lowdim}) describing the dynamics of the bump location $\theta(t)$ and the slow evolution of the nonlinearity $K(\theta,t)$, we focus on the limit $F(u) \to H(u - \kappa)$. We approximate $q(x,t)$ by summing the contributions from each of the $n+1$ trials. This yields
\begin{linenomath}
\begin{align}
q(x,t) \approx \sum_{j=1}^{n} {\mc A}_j(t) Q(x - \theta_q(t_j+T_C+T_D^n)) + {\mc A}_{n+1}(t) Q(x - \theta_q(t))  \label{origqanz}
\end{align}
\end{linenomath}
where the slowly evolving function ${\mc A}_n(t)$ defines the rising and falling kinetics of the facilitation variable originating in trial $n$:
\begin{linenomath}
\begin{align*}
\tau \dot{\mc A}_n(t) = \left\{ \begin{array}{ll} 1- {\mc A}_n(t) & t_n<t<t_n+T_C + T_D^n, \\ -{\mc A}(t) & t> t_n+ T_C + T_D^n, \end{array} \right.
\end{align*}
\end{linenomath}
increasing towards saturation (${\mc A}_n \to 1$) during the cue and delay-period $[t_n,t_n+T_C+T_D^n)$ and decaying afterward (${\mc A}_n \to 0$). The variable $\theta_q(t)$ describes the slow movement of the center-of-mass of the saturating portion of the facilitation variable $q(x,t)$ due to the drift of the neural activity $u(x,t)$ described by $\theta (t)$. However, since ${\mc A}_1(t) \ll {\mc A}_2(t) \ll \cdots \ll {\mc A}_{n}(t)$, we only keep the terms ${\mc A}_n(t)$ and ${\mc A}_{n+1}(t)$ in Eq.~(\ref{origqanz}). Furthermore, since ${\mc A}_n(t)$ becomes much smaller than ${\mc A}_{n+1}(t)$ for most times $t> t_{n+1}$ in trial $n+1$, we approximate $\theta_q(t_n+T_C+T_D^n) \approx \theta_n$. This provides intuition as to why it is sufficient to only consider the previous target rather than the response in trial $n$ as the variable influencing the bias in Papadimitriou et al. (2015)~\cite{papadimitriou15}. Therefore, we start with the following ansatz for the evolution of the facilitation variable during trial $n+1$:
\begin{linenomath}
\begin{align}
q(x,t) = {\mc A}_n(t) Q(x - \theta_n) + {\mc A}_{n+1}(t) Q(x - \theta_q(t)).  \label{panz}
\end{align}
\end{linenomath}
A bump centered at $\theta (t)$, $U(x - \theta (t))$, attracts the STF variable to the same location $q \to Q(x - \theta (t))$, but the dynamics of $q$ are much slower ($\tau \gg 1$). Thus, we model the evolution of $\theta_q(t)$ by linearizing the slow dynamics of Eq.~(\ref{bnet2}) about $(u,q) = (U(x-\theta(t)),Q(x - \theta (t)))+ (0,\phi(x,t))$ (with $|\phi| \ll 1$) to find
\begin{linenomath}
\begin{align}
\tau \dot{\phi}(x,t) = - \phi(x,t) - \beta F(U(x-\theta(t))) \phi (x,t).  \label{phieqn}
\end{align}
\end{linenomath}
The perturbation $\phi (x,t)$ describes the displacement of the variable $q$ away from its equilibrium position. We thus introduce the field $\Phi(x,t) = \int_{-180}^{180} w(x-y) \phi(y,t) F(U(y-\theta (t))) \d y$, which reduces Eq.~(\ref{phieqn}) to
\begin{linenomath}
\begin{align*}
\tau \dot{\Phi}(x,t) = - (1+ \beta) \Phi (x,t),
\end{align*}
\end{linenomath}
so separating variables $\Phi(x,t) = \bar{\Phi}(x) \e^{\lambda t}$ we see that perturbations of the facilitation variable's center-of-mass $\theta_q(t)$ away from $\theta (t)$ should relax at rate $\lambda_{\tau} = -(1+\beta)/\tau$.

Therefore, the slow evolution of the potential gradient function $K(\theta,t_s)$ in Eq.~(\ref{bumpsde}) can be described by integrating Eq.~(\ref{Kform}) using the ansatz Eq.~(\ref{panz}) for $q(x,t)$. Our low-dimensional system for the dynamics of the bump location $\theta (t)$ and leading order facilitation bump $\theta_q(t)$ during the delay-period of trial $n+1$ ($t \in [t_{n+1}+T_C,t_{n+1}+T_C+T_D^{n+1})$) is given by the set of non-autonomous stochastic differential equations:
\begin{linenomath}
\begin{align*}
\d \theta (t) &= - {\mc A}_n(t)  \frac{\d \bar{\mc U} (\theta - \theta_n)}{\d \theta} \d t - {\mc A}_{n+1}(t)  \frac{\d \bar{\mc U} (\theta - \theta_q(t))}{\d \theta} \d t + \sigma \d \xi (t), \\
\tau \dot{\theta}_q (t) &= - d(\theta_q(t)-\theta (t)),
\end{align*}
\end{linenomath}
where we have defined a parametrized time-invariant potential gradient $\frac{\d \bar{\mc U} (\theta - \theta')}{\d \theta}$ corresponding to the stationary profile of the facilitation variable centered at $\theta'$: $Q(x - \theta_n)$. For our specific choices of weight function and firing rate nonlinearity, we find the potential gradient is:
\begin{linenomath}
\begin{align*}
- \frac{\d \bar{\mc U} (\theta - \theta')}{\d \theta} =  \frac{\beta q_+}{2(1+\beta) \tan (a)} \left[ {\rm sign} (\theta-\theta' )(1- \cos(\theta - \theta' )) - \tan (a) \sin (\theta - \theta') \right],
\end{align*}
\end{linenomath}
and
\begin{linenomath}
\begin{align*}
d(\theta_q-\theta) = (1+ \beta) \left\{ \begin{array}{ll} \theta_q - \theta, & |\theta_q - \theta| \leq \pi \\  {\rm sign}(\theta_q)(2 \pi - |\theta_q - \theta| ), & |\theta_q - \theta|>\pi  \end{array} \right.
\end{align*}
\end{linenomath}
calculates the shorter difference on the periodic domain. As in our recurrent network, we use the parameters $\kappa = 0.1$; $q_+ = 2$; $\beta = 0.01$; and $\tau/\tau_u = 100$ to compare with network simulations in Fig. \ref{fig5_stats}.

\subsection*{Numerical simulations of the neuronal network model}

Numerical simulations of the recurrent network Eq.~(\ref{bumpnet}) were done in MATLAB using an Euler-Maruyama method with timestep $\d t = 0.1$ms and spatial step $\d x = 0.18^{\circ}$ with initial conditions generated randomly by starting $u(x,0) \equiv q(x,0) \equiv 0$ and allowing the system to evolve in response to the dynamic fluctuations for $t = 2$s prior to applying the sequence of stimuli $I(x,t)$ described for each numerical experiment in Figs. \ref{fig5_stats}, \ref{fig6_ensemble}, and \ref{fig7_compare}. Numerical simulations of Eq.~(\ref{lowdsys}) were also performed using an Euler-Maruyama method with timestep $\d t = 0.1$ms. The effects of the target $\theta_n$ on each trial $n$ were incorporated by holding $\theta(t) = \theta_n$ during the cue period $t \in [t_n, t_n + T_C)$. Otherwise, the dynamics were allowed to evolve as described.

\subsection*{Data Analysis}

MATLAB was used for statistical analysis of all numerical simulations. The bias effects in Fig. \ref{fig5_stats} were determined by identifying the centroid of the bump at the end of the delay-period. Means were computed across $10^5$ simulations each, and standard deviations were determined by taking the square root of the {\tt var} command applied to the vector of endpoints. Histograms in Fig. \ref{fig6_ensemble} were computed for $10^5$ simulations using the {\tt hist} and {\tt bar} commands applied to the vector of endpoints for each correlation condition. Bump positions were computed in Fig. \ref{fig7_compare} by determining the centroid of the bump at each timepoint, and $10^5$ simulations were then used to determine the standard deviation and variance plots (using {\tt var} again).



\section*{Author Contributions}

ZPK conceived the study, performed simulations and analysis, and wrote the paper.

\section*{Acknowledgments}

This work was funded by National Science Foundation grants NSF-DMS-1615737 and NSF-DMS-1517629. We thank Kre\v{s}imir Josi\'{c} and Brent Doiron for helpful conversations and comments on the manuscript.

\nolinenumbers

\bibliographystyle{unsrt}
\bibliography{repetition}

\end{document}